\documentclass[%
 reprint,
superscriptaddress,
 amsmath,amssymb,
 aps,
prb,
]{revtex4-2}

\usepackage{graphicx}
\usepackage{dcolumn}
\usepackage{bm}
\usepackage{hyperref}
\usepackage{siunitx}
\usepackage{braket}
\usepackage{placeins}
\usepackage{caption}

\begin{document}

\preprint{APS/123-QED}

\title{Magnon-mediated quantum gates for superconducting qubits}

\author{Martijn Dols}
\email{martijn.dols@rwth-aachen.de}
\affiliation{Institute for Theoretical Solid State Physics, RWTH Aachen University, 52074 Aachen, Germany}
\affiliation{Kavli Institute of Nanoscience, Delft University of Technology, 2628 CJ Delft, The Netherlands}

\author{Sanchar Sharma}
\affiliation{Laboratoire de Physique de l’\'{E}cole Normale Sup\'{e}rieure, ENS, Universit\'{e} PSL, CNRS, Sorbonne Universit\'{e}, Universit\'{e} de Paris, F-75005 Paris, France}

\author{Lenos Bechara}
\affiliation{Institute for Theoretical Solid State Physics, RWTH Aachen University, 52074 Aachen, Germany}

\author{Yaroslav M. Blanter}
\affiliation{Kavli Institute of Nanoscience, Delft University of Technology, 2628 CJ Delft, The Netherlands}

\author{Marios Kounalakis}
\email{marios.kounalakis@gmail.com}
\affiliation{Institute for Theoretical Solid State Physics, RWTH Aachen University, 52074 Aachen, Germany}
\affiliation{Kavli Institute of Nanoscience, Delft University of Technology, 2628 CJ Delft, The Netherlands}
\affiliation{Luxembourg Institute of Science \& Technology, 4422 Belvaux, Luxembourg}

\author{Silvia {Viola Kusminskiy}}
\email{kusminskiy@physik.rwth-aachen.de}
\affiliation{Institute for Theoretical Solid State Physics, RWTH Aachen University, 52074 Aachen, Germany}
\affiliation{Max Planck Institute for the Science of Light, Staudtstraße 2, 91058 Erlangen, Germany}

\date{\today}

\begin{abstract}
We propose a hybrid quantum system consisting of a magnetic particle inductively coupled to two superconducting transmon qubits, where qubit-qubit interactions are mediated via magnons. 
We show that the system can be tuned into three different regimes of effective qubit-qubit interactions, namely a transverse ($XX + YY$), a longitudinal ($ZZ$) and a non-trivial $ZX$ interaction.
In addition, we show that an enhanced coupling can be achieved by employing an ellipsoidal magnet, carrying anisotropic magnetic fluctuations. 
We propose a scheme for realizing two-qubit gates, and simulate their performance under realistic experimental conditions.
We find that iSWAP and CZ gates can be performed in this setup with an average fidelity  $\gtrsim 99 \% $ , while an iCNOT gate can be applied with an
average fidelity $\gtrsim 88 \%$.
Our proposed hybrid circuit architecture offers an alternative platform for realizing two-qubit gates between superconducting qubits and could be employed for constructing qubit networks using magnons as mediators.
\end{abstract}

\maketitle


\section{Introduction}\label{sec:intro}

Hybrid quantum systems provide a promising route towards practical applications by combining the advantages of different platforms for quantum information tasks~\cite{Kurizki2015Hybridsystems, Clerk2020Hybridsystems2}. For example, superconducting (SC) qubits make excellent processors for quantum computing~\cite{Vion2002SCQ_QI1,Clarke2008SCQ_QI2,Arute2019SCQ_QI3}. Quantum gates for SC qubits can be performed within several nanoseconds~\cite{Chow2010SingleQ,Barends2014SingleQ2_CZ,McKay2017SingleQ3}, with fidelities exceeding $99 \%$, as required for the error correction schemes of surface codes~\cite{Wang2011F_thresh1,Fowler2012F_thresh2,Fowler2012F_thresh3}. However, the dissipation rates of SC qubits make them impractical for long-term data storage, imposing the need for integration with more appropriate physical systems in order to construct quantum memories~\cite{Kurizki2015Hybridsystems, Clerk2020Hybridsystems2}.
Furthermore, SC circuits do not couple directly to optical photons, which is an important requirement for building quantum networks~\cite{Kurizki2015Hybridsystems,Clerk2020Hybridsystems2}.
Therefore, there is a practical need for bridging systems of diverse nature and functionality.

\begin{figure}
\includegraphics[width=\linewidth]{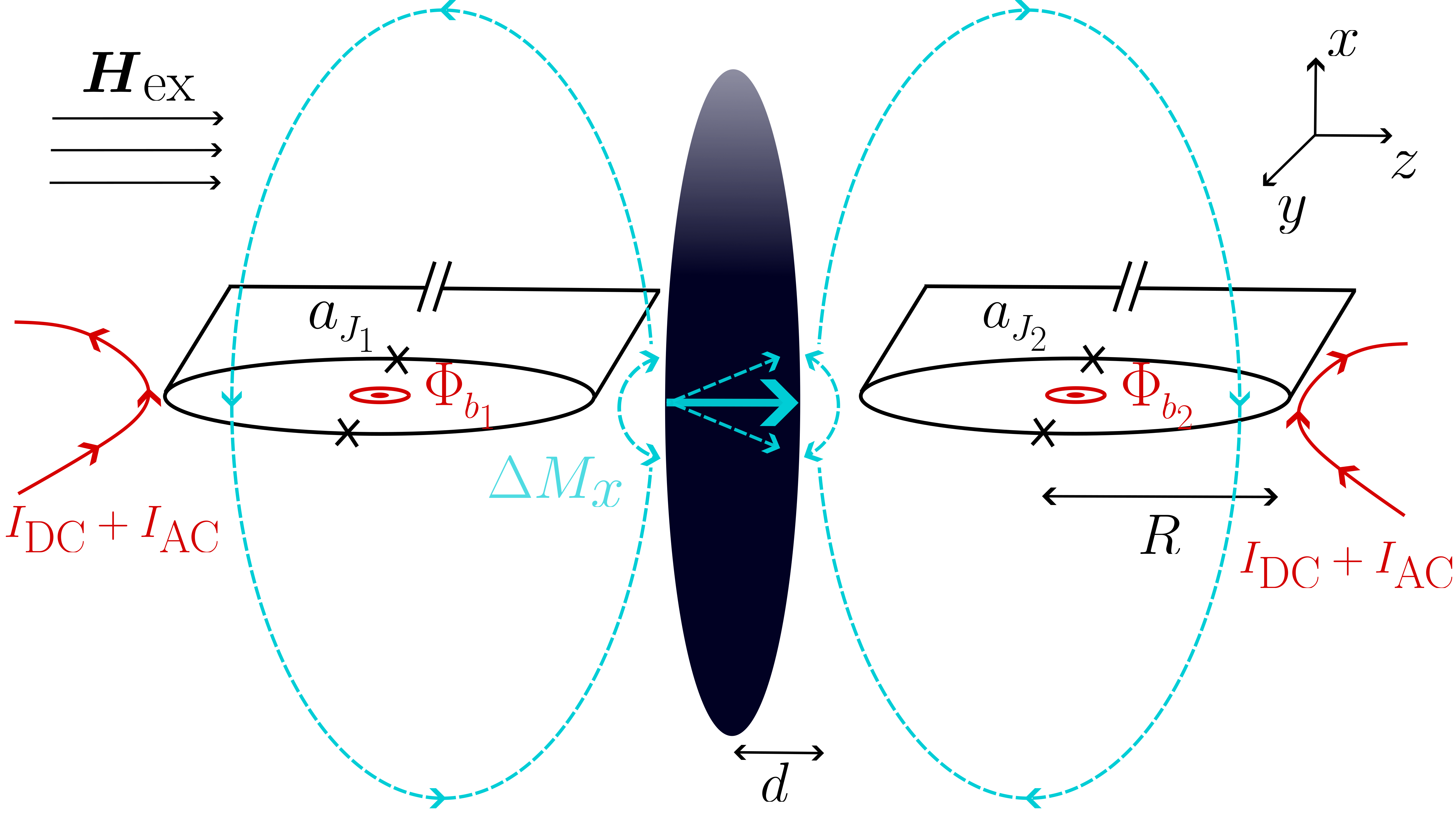}
\caption{\label{fig:setup} Proposed setup consisting of two transmon qubits inductively coupled via their SQUID loop to a YIG ellipsoid. The ground state of the magnetization is along the $z$ axis due to the external magnetic field $\boldsymbol{H}_{\text{ex}}=H_{0} \boldsymbol{e}_{z}$. The magnetic quantum fluctuations, $\Delta M_{x}$, of the YIG ellipsoid in the $x$ direction induce a flux in each SQUID loop, thereby modulating the inductance of both qubits and resulting in an effective qubit-qubit coupling mediated by magnons.
Two dedicated flux bias lines, carrying DC and AC current, are used to control the flux in each loop.}
\end{figure}

Magnons, the collective excitations of ordered spin systems, have shown promising properties to operate as such mediators~\cite{Chumak2018magnon_spintronics,Yuan2022magnon_QI, Babak2022cavmagnonics}, owing to their capability of coupling coherently to various excitations, e.g. optical photons~\cite{Osada2016opmag2,Liu2016opmag5,Kusminskiy2016opmag,Zhang2016opmag4,Haigh2018opmag3}, microwaves~\cite{Soykal2010micromag,Huebl2013micromag,Tabuchi2014micromag,Zhang2014micromag2,Goryachev2014micromag}, phonons~\cite{Weiler2012phononmag,Zhang2016phononmag,An2020phononmag,Potts2021phononmag,Schlitz2022phononmag2,Mueller2024phononmag}, and spins~\cite{Casola2018spinmag,Bertelli2020spinmag,Bejarano2024spinmag}.
These couplings can be further enhanced using magnetization squeezing~\cite{Kamra2020magnonsqueezing}, which can be implemented using anisotropically shaped magnetic structures~\cite{Sanchar_magnet}.
Moreover, by considering insulating magnetic materials, such as the paradigmatic Yttrium Iron Garnet (YIG), magnon dissipation channels can be minimized ~\cite{Cherepanov1993YIG}.
These properties suggest that magnons can be useful for mediating the coupling between different types of qubits, e.g. in spin-based or SC-based platforms.
The coupling of magnons to SC transmon qubits has been indeed experimentally demonstrated via mediating microwave cavities~\cite{Tabuchi2015transphotmag1, Lachance2020ransphotmag2, Wolski2020transphotmag3, Xu2023transphotmag4}.
Furthermore, it has been proposed that coherent coupling between a transmon qubit and a YIG sphere can also be achieved by the dipolar fields in free space, resulting in an interaction that can be tuned between a radiation-pressure and an exchange type~\cite{Marios_setup}. This interaction can, for example, be employed in order to control and entangle magnons in distant YIG spheres, using the qubit as a mediator and bypassing the need for microwave cavities~\cite{kounalakis2023engineering}.

In this work, we extend this framework by exploring the possibility of using magnons as mediators to couple SC qubits in free space. Various methods have been demonstrated to couple qubits to each other directly, including capacitive~\cite{Dewes2012capcoup2, Barends2013capcoup, Barends2014SingleQ2_CZ, Kandala2021CNOT} and inductive~\cite{You2005indcoup,Grajcar2006indcoup2,Niskanen2007indcoup3,Chen2014CZ2,kounalakis2018tuneable} coupling. Alternatively, a cavity bus or other qubits can be employed to mediate the coupling~\cite{majer2007cavitybus,McKay2016iSWAP,Roth2017iSWAP2,Xu2020CZ}.
These schemes focus on either direct coupling via circuit elements or indirect coupling via other circuit modes. Adding magnons to the list of possible mediators can open up new directions in qubit-qubit coupling schemes, e.g. by harnessing chiral coupling~\cite{Chen2019chirality2,Yu2020chirality1}. We show that, using magnons as virtual mediators of a tunable qubit-qubit interaction, a set of quantum gates can be realized with different degrees of fidelity which we characterize and optimize.
Specifically, we propose a hybrid quantum system consisting of two flux-tunable transmon qubits coupled via a magnet.
We consider a magnet with an anisotropic shape, which can enhance the coupling strength.
We find that, by appropriately choosing the SC qubit parameters, three different types of effective qubit-qubit interactions can be realized: $XX+YY$, $ZZ$, and $ZX$.
We use these to simulate three two-qubit gates, namely an iSWAP gate, a controlled-Z (CZ) gate and an iCNOT gate, respectively. Using experimentally realistic parameters for the setup, we obtain average gate fidelity values $\gtrsim 99 \%$ for the iSWAP and CZ gate, and $\gtrsim 88 \%$ for the iCNOT gate. The combination of any of these two-qubit gates with single-qubit rotations forms a universal gate set~\cite{Bremner2002Universal,Schuch2003iSWAP_uni}.

The remainder of this manuscript is structured as follows.
In Sec.~\ref{sec:modcoup} we review the theory of flux-tunable transmon qubits coupled to an anisotropic magnet leading to squeezing of the relevant quadrature of the magnetization fluctuations.
In Sec.~\ref{sec:gates} we derive the different regimes of magnon-mediated qubit-qubit interaction giving rise to the corresponding gates, and compare the performance of the dissipative system against the ideal case by defining a gate-fidelity measure. In Sec.~\ref{sec:con} we present the conclusions and discuss possibilities for improving the average gate fidelity values. Some details of the calculations have been relegated to the Appendix.

\section{Model and Coupling}\label{sec:modcoup}

We consider two flux-tunable SC transmon qubits coupled to a magnet of an anisotropic shape as depicted in Fig.~\ref{fig:setup}. A flux-tunable transmon qubit~\cite{Koch2007transmon} consists of
a superconducting quantum interference device (SQUID) made of an SC loop with two Josephson junctions with Josephson energies $E_{J}^{1}$ and
$E_{J}^{2}$, shunted
by a capacitor $C$ with charging energy $E_{C}=e^{2}/(2C)$.
An important parameter is the SQUID asymmetry, given by $a_{J}=|E_{J}^{1}-E_{J}^{2}|/E_{J}^{\Sigma}$, where $E_{J}^{\Sigma}=E_{J}^{1}+E_{J}^{2}$.
Applying current through a nearby external bias line induces a flux $\Phi_{b}$ into the SQUID loop. We define the reduced flux as $\varphi_{b}=\pi\Phi_{b}/\Phi_{0}$, where $\Phi_{0}=h/(2e)$ is the magnetic flux quantum.
The transmon Hamiltonian reads~\cite{Koch2007transmon}
\begin{equation}
\hat{H}_{Q}=4E_{C}\hat{n}^{2}-E_{J}^{\Sigma}S(\varphi_{b})\cos(\hat{\varphi}),\label{eq:H_Q}
\end{equation}
where $\hat{n}$ corresponds to the number of Cooper pairs participating in tunnelling, $S(\varphi_{b})=\sqrt{\cos^{2}(\varphi_{b})+a_{J}^{2}\sin^{2}(\varphi_{b})}$, and $\hat{\varphi}=\hat{\delta}-\arctan(a_{J}\tan(\varphi_{b}))$ is the SC phase difference.
We consider the transmon regime $E_{J}^{\Sigma}S(\varphi_{b})\gg E_{C}$, in which the qubit becomes insensitive to charge noise~\cite{Koch2007transmon}.
Expanding the cosine and introducing the annihilation and creation operators
$\hat{c}$ and $\hat{c}^{\dagger}$,
\begin{equation}
    \hat{n}=i\frac{\epsilon}{2}(\hat{c}^{\dagger}-\hat{c}), \; \; \hat{\varphi}=\frac{1}{\epsilon}(\hat{c}^{\dagger}+\hat{c})
\end{equation}
with $\epsilon^{4}=E_{J}^{\Sigma}S(\varphi_{b})/(2E_{C})$, one obtains~\cite{Koch2007transmon}
\begin{equation}
  \hat{H}_{T}=\hbar\omega_{q}\hat{c}^{\dagger}\hat{c}-\frac{E_{C}}{2}\hat{c}^{\dagger}\hat{c}^{\dagger}\hat{c}\hat{c},
\label{eq:H_T}
\end{equation}
where we defined the transmon frequency $\omega_{q} =\Bigl(\sqrt{8E_{C}E_{J}^{\Sigma}S(\varphi_{b})}-E_{C} \Bigr)/\hbar$.
The non-linearity of the second term in Eq.~(\ref{eq:H_T}) results in anharmonic energy levels, which is a necessary condition for the construction of a qubit~\cite{DiVincenzo}.

For the magnet we consider an ellipsoidal shape of dimensions $L_{x}\gg L_{y}=L_{z}$, where $L_{i}$ is the length of the semi axis of the magnet in the $i$-th direction (see  Fig.~\ref{fig:setup}) and $V_m = 4 \pi L_x L_z^2 / 3$ its volume.
An external homogeneous magnetic field is applied along the $z$ axis, $\boldsymbol{H}_{\text{ex}}=H_{0}\boldsymbol{e}_{z}$.  This field fulfills $H_{0}>M_{s}/2$
such that the classical ground state of the magnet is $\boldsymbol{M}=M_{s}\boldsymbol{e}_{z}$, where $M_{s}$ is the saturation magnetization.
The shape anisotropy favors fluctuations of the magnetization along the $x$ direction, implying that the quantum fluctuations of the ground state $\ket{0}_{m}$, denominated a \emph{squeezed vacuum}, are anisotropic~\cite{squeez}. In what follows we consider only the magnon excitations associated with the uniform precession of the spins, called Kittel magnons. Within the linear spin wave approximation they are described by 
\begin{equation}\label{eq:HM}
\hat{H}_{M}=\hbar\omega_{m}\hat{m}^{\dagger}\hat{m},
\end{equation}
where $\hat{m}^{(\dagger)}$ are the magnon annihilation (creation) operators operating on the squeezed vacuum such that $\hat{m}|0\rangle_m=0$, and the frequency is given by~\cite{Sanchar_magnet}
\begin{equation}
    \omega_{m}=\mu_{0}\gamma_{0} H_{0}\sqrt{1-\frac{M_{s}}{H_{0}}(3N_{T}-1)}\,. \label{def:omegam}
\end{equation}
Here, $\gamma_0$ ia the modulus of the gyromagnetic ratio and $N_T$ is a dimensionless factor of the demagnetization tensor which depends on the shape of the magnet~\cite{SpinWaves, Demagnetization}. For $L_{x}\gg L_{z}$ as considered in this work we have $N_{T}\approx\frac{1}{2}$~\cite{Demagnetization}. In the isotropic case $N_T=1/3$ the Kittel mode frequency for a spherical magnet, independent of demagnetization factors, is recovered.

Fluctuations of the magnetization, which are proportional to $1/\sqrt{V_{m}}$~\cite{Sanchar_magnet}, give rise to fluctuations of the magnetic dipole moment $\Delta \hat{\boldsymbol{\mu}}$, 
\begin{equation}
    \Delta \hat{\mu}_x = \mu_{\text{zpf}} e^r ( \hat{m} + \hat{m}^\dagger ), \quad \Delta \hat{\mu}_y = i \mu_{\text{zpf}} e^{-r}  (\hat{m} - \hat{m}^\dagger ).
\label{eq:mu_x}
\end{equation}
Here, $\mu_{\text{zpf}} = \hbar \gamma_0 \sqrt{ N_s /2 }$ is the value of the isotropic zero point fluctuations with $N_s=\rho_{s} V_{m}$ the total number of spins and $\rho_{s}$ the spin density. 
Magnets with volumes greater than $\SI{100}{nm^3}$ and typical YIG spin densities \cite{Tabuchi2015transphotmag1} obey $N_s \gg 1$. 
Since $\Delta \hat{\mu}_{z} / \mu_{\text{zpf}} \propto 1/N_{s}$, we can neglect the $z$ fluctuations of the magnetic moment~\cite{Rusconi2019geofactor}.
The factor $r$ is the \emph{squeezing parameter} and satisfies~\cite{Sanchar_magnet}
\begin{equation}
    e^{r}=\left(1-\frac{M_{s}}{H_{0}}(3N_{T}-1)\right)^{-\frac{1}{4}}.  
\end{equation}
In the limit $H_0 \rightarrow (3 N_T - 1 )M_s $, the magnon frequency $\omega_m$ vanishes, see Eq.~\eqref{def:omegam}, and the squeezing parameter $r$ diverges. In this case, $\Delta \hat{\mu}_{x}$ exponentially diverges with $r$, whereas $\Delta \hat{\mu}_{y}$ is suppressed. In practice, however, achievable cryogenic temperatures and the stability of the magnetically ordered ground state for these values of $H_0$ impose a lower bound on $\omega_m$, and therefore an upper bound on squeezing~\cite{Sanchar_magnet}.

We consider the magnet positioned between the qubits as shown in Fig.~\ref{fig:setup}.
The magnetic field due to the fluctuations $\Delta \hat{\boldsymbol{\mu}}$ induces a flux through the SQUID loop
\begin{equation}
    \Phi(\Delta\hat{\boldsymbol{\mu}}) = \frac{\mu_{0}}{4\pi} \sum_{i=x,y}I_{i}\Delta\hat{\mu}_{i},
\label{eq:phasecoupling}
\end{equation}
where $\mu_{0}$ is the magnetic constant and $I_{i}$ are geometrical factors which have the dimension of 1/length.
Due to the symmetry of the chosen setup one finds $I_y = 0$.
The geometrical factor $I_{x}$ is the largest for SQUID loops which are positioned at the $x=0$ plane and increases as $d$ (the minimal distance between the center of the magnet and the SQUID loop) decreases.
A lower bound on $d$ is imposed by the critical field that the SC wire can support~\cite{Rusconi2019geofactor}.
To determine $I_{x}$, we use the field of an ellipsoid~\cite{Chang1961Ellipsefield}; see details in App.~\ref{app:geo}.
We find $I_{x} = - \SI{0.1}{}/\SI{}{\micro m}$ for $L_{x}/L_{z} \approx 4$. The stray magnetic field of the magnet is about two orders of magnitude lower than the critical field of typical superconductors~\cite{Popinciuc2012Criticalfield}, as we elaborate in App~\ref{app:geo}.

The flux bias $\varphi_b$ can be controlled by the magnetic field generated by the wires carrying electric currents as shown in Fig.~\ref{fig:setup}. The reduced flux caused by $\Delta \hat{\boldsymbol{\mu}}$ is given by
\begin{equation}
  \varphi(\Delta\hat{\boldsymbol{\mu}}) = \frac{\mu_{0} I_{x} \mu_{\text{zpf}} e^r }{4 \Phi_0 } ( \hat{m}^\dagger + \hat{m} ).
\end{equation}
Replacing $\varphi_b \rightarrow \varphi_b +\varphi(\Delta\hat{\boldsymbol{\mu}})$ in Eq.~\eqref{eq:H_Q} and 
considering flux fluctuations much smaller than $\varphi_{b}$ gives the interaction Hamiltonian $\hat{H}_{\text{int}} = \hat{H}_{J} + \hat{H}_{g}$~\cite{Marios_setup}. The first term is a coherent exchange interaction between a qubit and the magnons $$\hat{H}_{J} = \hbar J(\hat{c}^{\dagger}\hat{m}+\hat{c}\hat{m}^{\dagger})$$ with the coupling constant
\begin{equation}
  J= - \frac{\mu_{0} I_{x} \mu_{\text{zpf}} a_{J} e^{r} }{4\Phi_{0}}\left(\frac{2E_{C}(E_{J}^{\Sigma})^{3}}{S(\varphi_{b})^{5}}\right)^{\frac{1}{4}},
\label{eq:J}
\end{equation}
while the second term is given by $$\hat{H}_{g}=\hbar g\hat{c}^{\dagger}\hat{c}(\hat{m}^{\dagger} + \hat{m})$$ with the coupling strength
\begin{equation}
  g= - \frac{\mu_{0} I_{x} \mu_{\text{zpf}} e^r }{8\Phi_{0}}\left(\frac{2E_{C}E_{J}^{\Sigma}}{S(\varphi_{b})^{3}}\right)^{\frac{1}{2}}\sin(2\varphi_{b})(1-a_{J}^{2}).
\label{eq:g}
\end{equation}
This second interaction term resembles an optical photon-magnon coupling~\cite{Kusminskiy2016opmag} or radiation pressure in optomechanical systems~\cite{RP}. Note that, because of the enhanced fluctuations of the magnetic moment $\Delta \hat{\mu}_{x}$ due to squeezing, the coupling strengths $J$ and $g$ are enhanced by $e^r$.

The total Hamiltonian of the system with two SC qubits therefore reads 
\begin{equation}
\hat{H}_{\rm{tot}}=\hat{H}_{0} + \hat{H}_{\text{int}}\,,  
\label{eq:H_tot}
\end{equation}
where
\begin{equation}
    \hat{H}_{0} = \hat{H}_{M} + \sum_{i=1,2} \hat{H}_{T}^{i},
\label{eq:H_0}
\end{equation}
with $\hat{H}_{M}$ the magnon Hamiltonian given by Eq.~\eqref{eq:HM}, and
\begin{equation}
    \hat{H}_{T}^{i} = \hbar \omega_{q_{i}} \hat{c}^{\dagger}_{i}\hat{c}_{i} - \frac{E_{C}}{2}\hat{c}^{\dagger}_{i}\hat{c}^{\dagger}_{i}\hat{c}_{i}\hat{c}_{i}
\end{equation} 
is the Hamiltonian for each SC qubit labeled by $i\in\{1,2\}$. The interaction term between the SC qubits and the magnon mode is given by
\begin{equation}
    \hat{H}_{\text{int}} = \sum_{i=1,2} \hat{H}_{J}^{i} + \hat{H}_{g}^{i},
\label{eq:H_int}
\end{equation}
with 
\begin{equation}
\hat{H}_{J}^{i} = \hbar J_{i} (\hat{c}^{\dagger}_{i}\hat{m}+\hat{c}_{i}\hat{m}^{\dagger})   
\end{equation}
and
\begin{equation}
\hat{H}_{g}^{i}=\hbar g_{i}\hat{c}^{\dagger}_{i}\hat{c}_{i}(\hat{m}^{\dagger}+\hat{m}),
\end{equation}
where the coupling constants of the magnons to each qubit can be tuned independently and are given by Eqs.~\eqref{eq:J} and~\eqref{eq:g}.

\section{Quantum gates}\label{sec:gates}

The Hamiltonian $\hat{H}_{\rm{tot}}$ can be brought into the form of an effective qubit-qubit interaction up to second order in the coupling constants $J_i$ and $g_i$ by performing a Schrieffer-Wolff (SW) transformation, as we show in Appendix~\ref{app:SW}. The effective interaction allows us to identify the coupling parameters that are required in order to realize different gates by appropriately tuning the coupling constants. These can be tuned by controlling the SQUID asymmetry $a_{J}$ (which is a design parameter) for each qubit, 
and the reduced flux $\varphi_{b}$. We identify two limiting cases: (i) for symmetric SQUIDs, $a_J = 0$, where the coupling strength
$J=0$, following Eq.~(\ref{eq:J}), and (ii)
for a highly asymmetric SQUID with $a_J \rightarrow 1$ and a value of the reduced flux $\varphi_{b} = \pi/2$, where the coupling constant $g$ vanishes according to Eq.~(\ref{eq:g}). In Appendix~\ref{app:SW} we show that combinations of these limiting cases give rise to the three gates studied in this section: iSWAP, CZ and iCNOT. In what follows we characterize the performance of the gate generated by the Hamiltonian $\hat{H}_{\rm{tot}}$ of Eq.~(\ref{eq:H_tot}) for the parameter regime of each gate (which we denote by $\hat{H}_{\text{gate}}$) and in the presence of dissipation, compared to the ideal gate associated with the corresponding effective qubit-qubit Hamiltonian. 

In order to take into account the dissipative evolution of the system we use a Liouvillian description. Depending on the gate, we apply a specific Hamiltonian $\hat{H}_{\text{gate}}$ and hence a specific Liouvillian $\mathcal{L}_{\text{gate}}$. The time evolution of the density matrix of the composite system $\hat{\rho}_{c}$, describing both qubits and the magnon field, can be found by solving the Lindblad master equation $\frac{\text{d}}{\text{d} t} \hat{\rho}_{\text{c}} (t)= \mathcal{L}_{\text{gate}}[\hat{\rho}_{\text{c}}(t)]$, where
\begin{equation}
\begin{aligned}
    \mathcal{L}_{\text{gate}}[\hat{\rho}_{\text{c}}]& =-\frac{i}{\hbar}\left[\hat{H}_{\text{gate}}, \hat{\rho}_{\text{c}} \right] \\
    & \quad +\sum_{n=1}^{6} \left( \hat{L}_{n} \hat{\rho}_{\text{c}} \hat{L}^{\dagger}_{n}
    - \frac{1}{2} \left \{ \hat{L}_{n}^{\dagger}\hat{L}_{n}, \hat{\rho}_{\text{c}} \right \}  \right).
\end{aligned}
\label{eq:lindblad}
\end{equation}
Here, $\hat{L}_{n}$ are the Lindblad operators. Magnon damping is taken into account by $\hat{L}_{1}=\sqrt{\kappa(1+n_{\text{th}})}\hat{m}$,
where $\kappa$ is the magnon linewidth and the bosonic expectation number $n_{\text{th}}=(\exp [\hbar\omega_{m} / (k_{B}\mathcal{T})] - 1)^{-1}$, with $k_B$ as the Boltzmann factor and $\mathcal{T}$ as the temperature. The linewidth is known from experiments to have the form $\kappa=\alpha_{G}\omega_{m} + \tilde{\kappa}$ \cite{damping2,damping3,inhom}, where $\alpha_G$ is the Gilbert damping and $\tilde{\kappa}$ is the inhomogeneous damping.
The thermal excitation of magnons is given by $\hat{L}_{2}=\sqrt{\kappa n_{\text{th}}}\hat{m}^{\dagger}$.
For the qubits we include the decay of the transmons with $\hat{L}_{3}=\sqrt{1/T_1}\hat{c}_{1}$ and $\hat{L}_{4}=\sqrt{1/T_1}\hat{c}_{2}$, where $T_1$ is the qubit lifetime, 
and pure dephasing terms $\hat{L}_{5}=\sqrt{1/T_{\phi}}\hat{c}^{\dagger}_{1}\hat{c}_{1}$ and $\hat{L}_{6}=\sqrt{1/T_{\phi}}\hat{c}^{\dagger}_{2}\hat{c}_{2}$, where $T_{\phi}$ is the dephasing time.

We assume for simplicity that initially the magnons are prepared in the vacuum state, and we describe the two-qubit space by the density matrix $\hat{\rho}$. This assumption does not affect significantly our results as long as the initial thermal occupation is low, as we discuss in Sec.~\ref{sec:CZ}. We propagate the initial state until an appropriately chosen gate time $T_{\text{gate}}$ at which the desired gate is applied. After the time propagation, the magnons are traced-out, since we are solely interested in the qubit dynamics. In summary, we consider the following quantum channel,
\begin{equation}
  \mathcal{E}_{\text{gate}}[\hat{\rho}] = \text{Tr}_{m} \left [ e^{ \mathcal{L}_{\text{gate}} T_{\text{gate}} } \left[\hat{\rho} \otimes \ket{0}\bra{0}_{m} \right]  \right].
\label{eq:gate}
\end{equation}
To quantify how well $\mathcal{E}_{\text{gate}}$ simulates a given gate $\hat{U}_{\text{gate}}$, we determine the average gate fidelity given by $\bar{F}(\mathcal{E}_{\rm gate}, \hat{U}_{\rm gate})$ where $\bar{F}$ is defined as \cite{fidelity}
\begin{equation}
  \bar{F} \bigl(\mathcal{E},\hat{U} \bigl) = \int \text{d}\psi \bra{\psi} \hat{U}^{\dagger} \mathcal{E}[\ket{\psi} \bra{\psi}] \hat{U} \ket{\psi}. \label{def:fid}
\end{equation}
Here, one integrates over the uniform measure $\text{d}\psi$, which is normalized such that $\int \text{d}\psi = 1$. This integral can be simplified to a finite summation over a unitary basis~\cite{fidelity}, as we elaborate in Appendix~\ref{app:fid}.

\begin{figure}
\includegraphics[width=\linewidth]{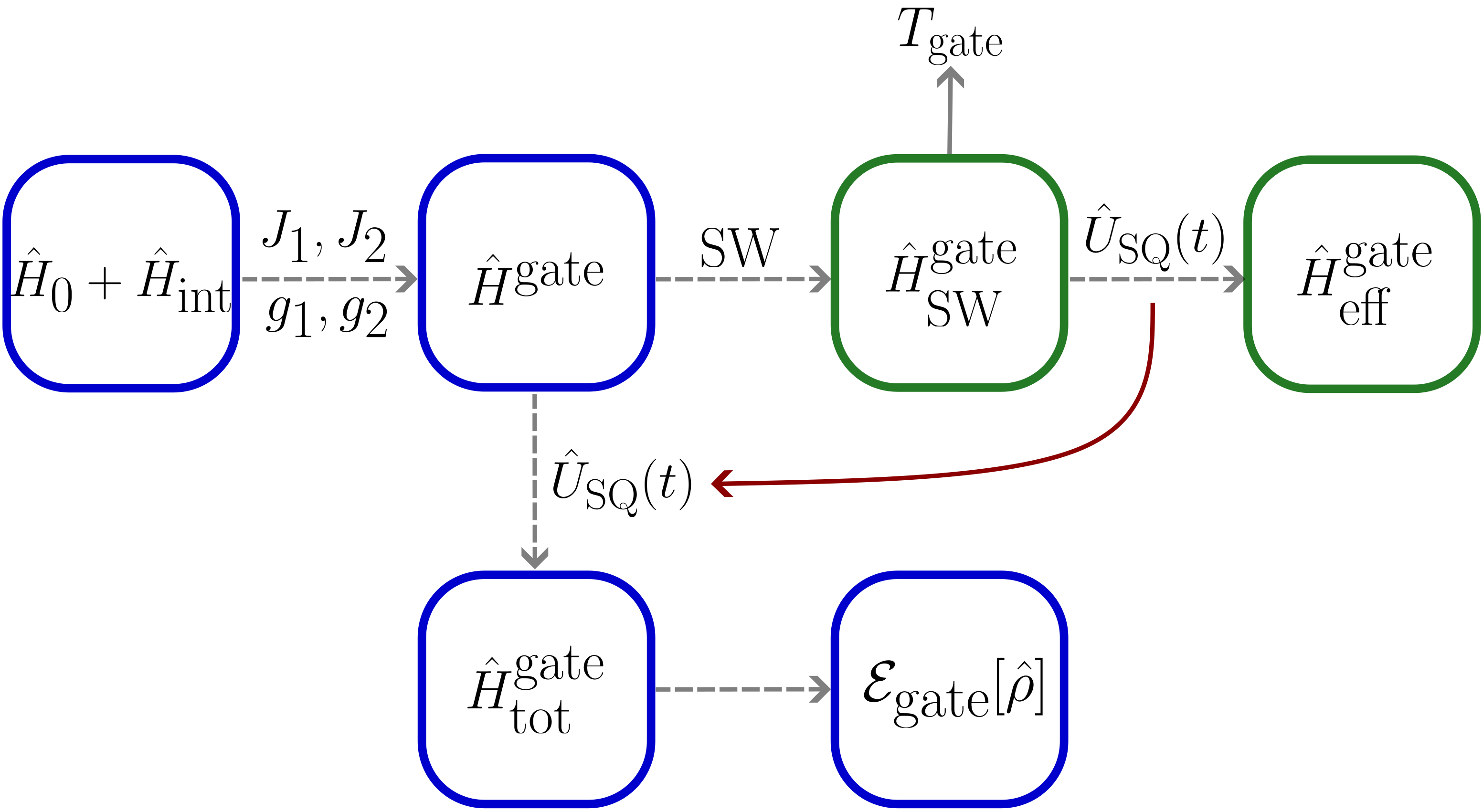}
\caption{\label{fig:scheme} Visual summary of the method used to obtain the quantum channel $\mathcal{E}_{\text{gate}}[\hat{\rho}]$. A given set of coupling constants $J_{1}$, $J_{2}$, $g_{1}$ and $g_{2}$ yields the total Hamiltonian for a particular gate, $\hat{H}^{\text{gate}}$. With a SW transformation we obtain an approximate qubit-qubit Hamiltonian $\hat{H}_{\text{SW}}^{\text{gate}}$, where the effective coupling strength dictates the gate time $T_{\text{gate}}$. We cancel single-qubit rotations with a unitary transformation $\hat{U}_{\text{SQ}}(t)$ to obtain the effective Hamiltonian $\hat{H}_{\text{eff}}^{\text{gate}}$. We apply this unitary transformation to the initial Hamiltonian as well to find $\hat{H}_{\text{tot}}^{\text{gate}}$. We use this final Hamiltonian to construct the quantum channel, which we use in our simulations. 
For the system Hamiltonians indicated by blue boxes, all quantum levels are included in the analysis, whereas green boxes indicate that the two-level approximation is applied.}
\end{figure}

One can find the SW transformed Hamiltonian
with general qubit levels in Appendix~\ref{app:SW}. 
In order to find the gate times which come from the effective qubit-qubit coupling and to identify the frame in which the gate is performed, we limit the qubits of these effective Hamiltonians to their energetically lowest two levels in the next sections.
One can recognize Hamiltonians in the two-level approximation by the usage of the Pauli operators:
$\hat{\sigma}^{+}=\ket{1}\bra{0}$, $\hat{\sigma}^{-}=\ket{0}\bra{1}$, and $\hat{\sigma}^{z} = \ket{1}\bra{1}$.
In simulations we use Hamiltonians which do not involve truncated qubit levels and are written in terms of the ladder operators $\hat{c}_{i}$ and $\hat{c}_{i}^{\dagger}$. We summarize the strategy applied to evaluate the different gates in Fig.~\ref{fig:scheme}.

\subsection{iSWAP}\label{sec:iSWAP}

For highly asymmetric SQUIDs with $a_{J_{1}},a_{J_{2}}\rightarrow 1$ and reduced flux $\varphi_{b_{1}},\varphi_{b_{2}}=\pi/2$
we have $g_{1},g_{2}\rightarrow0$. We obtain the effective qubit-qubit interaction for these parameters from $\hat{H}_{\rm{tot}}$ as detailed in Appendix~\ref{sec:SW_iSWAP}. We find
\begin{equation}
\begin{aligned}
  \hat{H}_{\text{SW}}^{\text{iSWAP}}&= \hbar\left(\omega_{m} -\sum_{i=1,2} \frac{J_{i}^{2}}{\omega_{q_{i}} - \omega_{m}} \right) \hat{m}^{\dagger}\hat{m} \\ & \quad +\sum_{i=1,2} \hbar\left(\omega_{q_{i}}+\frac{J_{i}^{2}}{\omega_{q_{i}}-\omega_{m}}\right)\hat{\sigma}^{z}_{i}
  \\ & \quad +\hbar g_{S}(\hat{\sigma}_{1}^{+}\hat{\sigma}_{2}^{-} + \hat{\sigma}_{1}^{-}\hat{\sigma}_{2}^{+}),
\end{aligned}
  \label{eq:H_SW_trans}
\end{equation}
with the effective qubit-qubit coupling constant
\begin{equation}
g_{S}=\frac{J_{1}J_{2}}{2}\left(\frac{1}{\omega_{q_{1}}-\omega_{m}}+\frac{1}{\omega_{q_{2}}-\omega_{m}}\right).
\label{eq:g_xy}
\end{equation}
We note that besides the exchange-like induced qubit-qubit coupling term, the interaction induces a Stark shift of the frequencies $\omega_{m}$ and $\omega_{q_i}$. This transformation is valid for $J_{i} \ll \omega_{q_{i}}-\omega_{m}$, i.e. in the dispersive regime.

Time propagation of the coupling term of the Hamiltonian of Eq.~(\ref{eq:H_SW_trans}) for a time $T_{S} = \pi/(2|g_{S}|)$ gives an $\text{iSWAP}$ gate: $\exp\bigl(- i g_{S} T_{S} (\hat{\sigma}_{1}^{+}\hat{\sigma}_{2}^{-} + \hat{\sigma}_{1}^{-}\hat{\sigma}_{2}^{+})\bigr ) = \hat{U}_{\text{iSWAP}}$, where we defined
\begin{equation}
\hat{U}_{\text{iSWAP}} = \frac{1}{2} (\hat{I}\otimes \hat{I} \mp i \hat{\sigma}_x \otimes \hat{\sigma}_x \mp i \hat{\sigma}_y \otimes \hat{\sigma}_y + \hat{\sigma}_z \otimes \hat{\sigma}_z ).
\label{eq:iSWAP}
\end{equation}
Here, $\hat{\sigma}_{x}$, $\hat{\sigma}_{y}$ and $\hat{\sigma}_{z}$ are the Pauli matrices.
This gate swaps $\ket{01} \leftrightarrow \ket{10}$ while adding a phase $\mp i$ and leaves the symmetrical states $\ket{00}$ and $\ket{11}$ untouched. The sign of the phase $\mp i$ corresponds to $g_{S} \gtrless 0$.

\begin{figure*}[t]
\includegraphics[width=\linewidth]{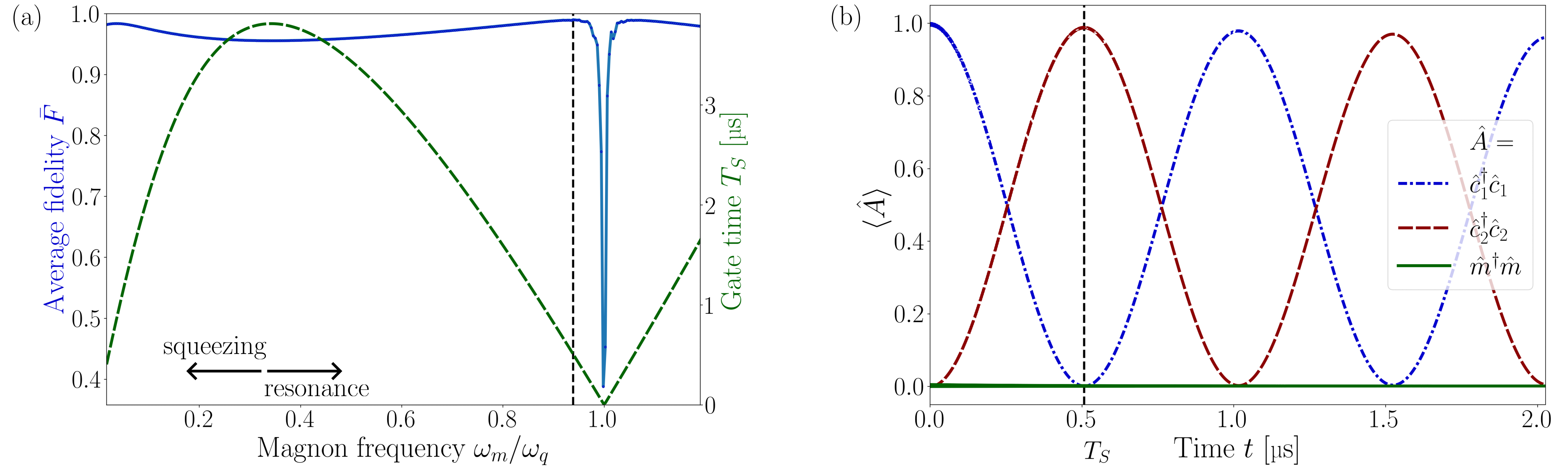}
\caption{\label{fig:iSWAP} (a) Average fidelity $\bar{F}$ (blue, solid) and gate time $T_{S}$ (green, dashed) for the iSWAP gate as a function of the magnon frequency $\omega_{m}$ in units of the SC qubit frequencies $\omega_{q}=\omega_{q_1}=\omega_{q_2}$. $T_{S}$ drops at low frequencies due to the squeezing enhancement of the qubit-qubit coupling strength $g_{S}$, and approaches $0$ for $\omega_{m}/\omega_{q}\rightarrow 1$ due to the resonant enhancement of $g_{S}$. Undesired magnonic excitations near resonance and thermal excitations at low frequencies cause the average fidelity to drop.
The maximum fidelity is $\bar{F}=99.00\%$ at $\omega_m/\omega_q = 0.94$ (vertical dashed line).
(b) Qubit and magnon dynamics following the evolution with the Hamiltonian of Eq.~(\ref{eq:H_tot_trans}) in the presence of dissipation at the optimal magnon frequency $\omega_m = 0.94 \, \omega_q$. The input state is $\ket{\psi} = \ket{10}$.
The occupation of both qubits (blue, dash-dotted and red, dashed) and magnons (green, solid) as a function of time $t$ are shown.
In this regime, the magnon occupation remains $\sim 0$ at all times. The gate time $T_S = \SI{0,5}{\micro s}$ is indicated with a vertical dashed line, at which the target state equals $\hat{U}_{\text{iSWAP}}\ket{\psi} = -i\ket{01}$. Parameters: $L_x = \SI{16}{\micro m}$, $L_z = \SI{3.9}{\micro m}$, $R=\SI{25}{\micro m}$, $d - L_{z} = \SI{10}{nm}$, $\alpha_G = 10^{-4}$, $\tilde{\kappa}/(2\pi)=\SI{0.1}{MHz}$ \cite{damping2,damping3,inhom} and $\mathcal{T} = \SI{10}{mK}$, $T_{1} = \SI{100}{\micro s}$ and $T_{\phi} = \SI{100}{\micro s}$ \cite{dissi_qubit,lifetime_qubit,coherence_qubit}.}
\end{figure*}

The second term of Eq.~(\ref{eq:H_SW_trans}) causes the qubits to undergo rotations on top of the interaction. We assume for simplicity identical qubits, so that $J=J_1=J_2$ and $\omega_q = \omega_{q_{1}} = \omega_{q_{2}}$. A unitary transformation with
\begin{equation}
    \hat{U}_{\text{SQ}}(t)=\exp\biggl[it\biggl(\omega_{q}+\frac{J^{2}}{\omega_{q}-\omega_{m}}\biggr)\biggl(\hat{m}^{\dagger}\hat{m}+\sum_{i=1,2}\hat{\sigma}_{i}^{z}\biggr)\biggr]
\label{eq:U_trans}
\end{equation}
cancels these phase rotations yielding
\begin{equation}
\begin{aligned}
  \hat{H}_{\text{eff}}^{\text{iSWAP}} & =\hat{U}_{\text{SQ}}\hat{H}_{\text{SW}}^{\text{iSWAP}}\hat{U}_{\text{SQ}}^{\dagger}+i\hbar\frac{\text{d} \hat{U}_{\text{SQ}}}{\text{d} t}\hat{U}_{\text{SQ}}^{\dagger}\\
 & =\hbar\left(\omega_{m} - \omega_{q} - \frac{3 J^2}{\omega_{q} - \omega_{m}} \right)\hat{m}^{\dagger}\hat{m} \\ & \quad + \hbar g_{S}(\hat{\sigma}_{1}^{+}\hat{\sigma}_{2}^{-} + \hat{\sigma}_{1}^{-}\hat{\sigma}_{2}^{+})\,.
\end{aligned}
\label{eq:H_eff_trans}
\end{equation}
In order to cancel the single-qubit rotations at the level of the original Hamiltonian, we apply this transformation to $\hat{H}_{\rm{tot}}$, which yields
\begin{equation}
\begin{aligned}
 \hat{H}_{\text{tot}}^{\text{iSWAP}}&=\hbar\left(\omega_{m}-\omega_{q}-\frac{J^{2}}{\omega_{q}-\omega_{m}}\right)\hat{m}^{\dagger}\hat{m} \\ & \quad - \sum_{i=1,2} \left(\frac{\hbar J^{2}}{\omega_{q}-\omega_{m}} \hat{c}_{i}^{\dagger}\hat{c}_{i} +\frac{E_{C}}{2}\hat{c}_{i}^{\dagger}\hat{c}_{i}^{\dagger}\hat{c}_{i}\hat{c}_{i}\right) \\ & \quad + \sum_{i=1,2} \hbar J(\hat{c}_{i}^{\dagger}\hat{m}+\hat{c}_{i}\hat{m}^{\dagger}).
\end{aligned}
\label{eq:H_tot_trans}
\end{equation}
We use this Hamiltonian to obtain the channel $\mathcal{E}_{\text{iSWAP}}[\hat{\rho}]$ and compare it to the ideal iSWAP gate given by Eq.~(\ref{eq:iSWAP}), by computing the average gate fidelity as a function of the magnon frequency, see Eq.~\eqref{def:fid}. The result is shown in Fig.~\ref{fig:iSWAP}(a).

We see that the average fidelity is a non-monotonic function of the magnon frequency, which is a consequence of two competing factors affecting the effective coupling strength $g_{S}$ and therefore the gate time $T_{S}$. Minimizing the gate time is important in order to minimize the effect of dissipation and hence to improve the gate fidelity. On the one hand, Eq.~(\ref{eq:g_xy}) shows that $g_{S}$ is enhanced as the magnon frequency approaches the resonance condition $\omega_m=\omega_q$. This is reflected in Fig.~\ref{fig:iSWAP}(a), where the gate time $T_{S}$ goes to zero when approaching $\omega_{m}/\omega_{q}=1$. Note however that, as this resonance is approached, the SW transformation used to obtain the ideal gate breaks down (signaled by a diverging coupling $g_{S}$) so that the effective, ideal-gate Hamiltonian is not a good approximation to the total one at this point. Physically, a resonant magnon-qubit coupling causes real magnonic excitations (as opposed to the virtual ones in the off-resonant case) which adversely affect the gate fidelity. This causes the rapid decrease of the average gate fidelity as the resonance is approached.
On the other hand, the coupling strength is enhanced by magnon squeezing. As Eq.~(\ref{eq:g_xy}) shows, the coupling is proportional to $J_{1}J_{2}$ and hence $g_{S}\propto e^{2r}$. In order to have $e^{r} \gg 1$ the magnon frequencies are required to be small $\omega_m \ll \gamma\mu_{0}M_{s}$, which corresponds to the far left side of Fig.~\ref{fig:iSWAP}(a). At such low frequencies, however, magnonic thermal excitations are important, decreasing the gate fidelity. The effect of squeezing around the qubit frequency region is generally negligible due to $e^{r} \approx 1$. This competition of effects gives rise to the maximum of the average gate fidelity at low frequencies.

For the results shown in Fig.~\ref{fig:iSWAP} we set $a_{J} = 0.9$ for both qubits, considering fabrication constraints~\cite{asymmetry_parameter} and in order to maintain the frequency tunability~\cite{Marios_setup}. The reduced flux of both qubits is set to $\varphi_b = \pi/2$ such that $g_{1}=g_{2}=0$, in agreement with the considered iSWAP regime. We assume typical transmon energies $E_C/h=\SI{150}{MHz}$ and $E_{J}^{\Sigma}/h = \SI{35}{GHz}$, such that the qubit frequency is $\omega_{q}/(2\pi) = \SI{6.0}{GHz}$~\cite{transmon_energy}.
In the simulations we use Fock spaces with size 3 for the qubits and 4 for the magnons~\cite{QuTip}.
A maximal fidelity of $\bar{F}=99.00\%$ is obtained at $\omega_m/\omega_{q} = \SI{0.94}{}$. For this magnon frequency the effective qubit-qubit coupling is $g_{S}/\omega_{q} = 8.2\times10^{-5}$ corresponding to $g_{S}/(2\pi) = \SI{0.49}{MHz}$ for the parameters used.

In Fig.~\ref{fig:iSWAP}(b) we show the dynamics of the system including dissipation for a given input state, $\ket{10}$ and for a magnon frequency tuned to the optimal average fidelity. From $t=0$ to $t=T_S$, which corresponds to the dashed line, the swap takes place. We see that the excitation of the first qubit is transferred to the second qubit, as we expect from $\hat{U}_{\text{iSWAP}} \ket{10} = -i \ket{01}$, whereas the magnon occupation remains close to zero as expected for virtual transitions.

In order to turn off the qubit-qubit coupling once the gate has been realized at the gate time $T_{S}$, the interaction term in Eq.~(\ref{eq:H_SW_trans}) can be made off resonant by detuning the qubits. This can be achieved by varying the reduced flux bias $\varphi_{b}$. Simulations show that changing the reduced flux of one qubit to $\varphi_{b} = \pi /3$ while keeping the other at $\varphi_{b} = \pi /2$ is sufficient.

The configuration described above can also be used to construct a $\sqrt{\text{iSWAP}}$ gate by choosing a gate time $T_{S}/2$.
This gate can be used to create Bell-like states, such as $\ket{01}-i\ket{10}$~\cite{sqrtSWAP}.
In our setup, using the same parameters as in Fig.~\ref{fig:iSWAP}, we find that such a state can be prepared with an average gate fidelity $\bar{F} = 99.54 \%$.

\subsection{CZ}\label{sec:CZ}

In the case of symmetric SQUIDs with $a_{J_{1}},a_{J_{2}}\rightarrow0$, we have $J_{1},J_{2}\rightarrow0$. In this case we obtain the following effective interaction Hamiltonian (see App.~\ref{sec:SW_CZ})
\begin{equation}
      \hat{H}_{\text{SW}}^{\text{CZ}} = \hbar\omega_{m} \hat{m}^{\dagger}\hat{m} + \sum_{i=1,2}\hbar\left(\omega_{q_{i}} - \frac{g_{i}^{2}}{\omega_{m}}\right)\hat{\sigma}_{i}^{z} -\hbar g_{Z}\hat{\sigma}_{1}^{z}\hat{\sigma}_{2}^{z}
\label{eq:H_SW_long}
\end{equation}
where the effective coupling strength is given by
\begin{equation}
    g_{Z}=\frac{2 g_{1} g_{2}}{\omega_{m}}.
\label{eq:g_zz}
\end{equation}
This transformation is valid for $g_{i} \ll \omega_{m}$. At $T_{Z}=\pi/g_{Z}$, the coupling term in Eq.~(\ref{eq:H_SW_long}) results in a CZ gate, since $\exp\bigl(ig_{Z}T_{Z}\hat{\sigma}_{1}^{z}\hat{\sigma}_{2}^{z}\bigr) = \hat{U}_{\text{CZ}}$, where we defined
\begin{equation}
  \hat{U}_{CZ} = \ket{0}\bra{0}\otimes \hat{I} + \ket{1}\bra{1}\otimes \hat{\sigma}_{z}.
\label{eq:CZ}
\end{equation}
If either of the qubits is excited, a Pauli gate $\hat{\sigma}_z$ is applied on the target qubit.

\begin{figure*}[t]
\includegraphics[width=\textwidth]{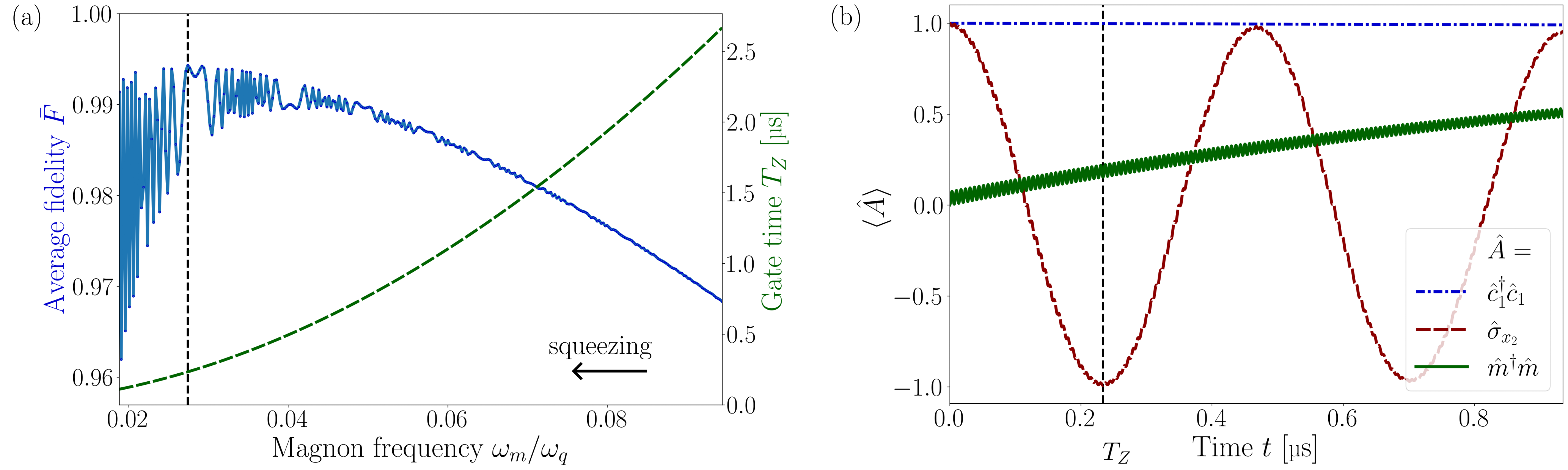}
\caption{\label{fig:CZ} (a) Average fidelity $\bar{F}$ (blue, solid) and gate time $T_{Z}$ (green, dashed) of the CZ gate as a function of the magnon frequency $\omega_{m}$ in units of the qubit frequencies $\omega_{q}$. A competition between squeezing enhancement of the coupling $g_{Z}$ and thermal occupation at low frequencies results in a maximum fidelity $\bar{F}=99.43\%$ at $\omega_m /\omega_{q} = 0.027$ (vertical dashed line).
The oscillatory behavior for low magnon frequencies is due to a resonant enhancement of the parametric interaction between magnons and qubits, causing an oscillating magnon occupation, see also green solid line in (b).
(b) Demonstration of the proposed CZ gate for a given input state ($\ket{1+}$) at the optimal magnon frequency $\omega_m = 0.027 \, \omega_{q}$. 
We plot the occupation of qubit 1 (blue, dash-dotted) and of the magnons (green, solid) as a function of time.
In addition, we plot the evolution of the expectation value of the Pauli $\hat{\sigma}_x$ operator for the second qubit (red, dashed).
The target state, achieved at $T_{Z}=\SI{0.23}{\micro s}$ (vertical dashed line), equals $\hat{U}_{\text{CZ}}\ket{1+} = \ket{1-}$. Parameters as in Fig.~\ref{fig:iSWAP}.}
\end{figure*}

The second term of the transformed Hamiltonian (\ref{eq:H_SW_long}) causes qubit rotations regardless of the state of the control qubit.
We cancel these with the unitary transformation
\begin{equation}
\hat{U}_{\text{SQ}}(t)=\exp\biggl[it\biggl(\, \sum_{i=1,2}\biggl(\omega_{q_{i}} - \frac{g_{i}^{2}}{\omega_{m}} \biggl)\hat{\sigma}_{i}^{z} \biggr)\biggr].    
\end{equation}
This yields
\begin{equation}    
    \hat{H}_{\text{eff}}^{\text{CZ}} = \hbar \omega_{m} \hat{m}^{\dagger}\hat{m}-\hbar g_{Z} \hat{\sigma}_{1}^{z} \hat{\sigma}_{2}^{z}
\label{eq:H_eff_long}
\end{equation}
for the effective Hamiltonian and 
\begin{equation}
\begin{aligned}
        \hat{H}_{\text{tot}}^{\text{CZ}} & = \hbar\omega_{m}\hat{m}^{\dagger}\hat{m} + \sum_{i=1,2}\frac{\hbar g_{i}^{2}}{\omega_{m}}\hat{c}_{i}^{\dagger}\hat{c}_{i} - \frac{E_{C}}{2}\hat{c}_{i}^{\dagger}\hat{c}_{i}^{\dagger}\hat{c}_{i}\hat{c}_{i} \\ & +\sum_{i=1,2}\hbar g_{i}\hat{c}_{i}^{\dagger}\hat{c}_{i}(\hat{m}^{\dagger}+\hat{m})
\end{aligned}
\label{eq:H_tot_long}
\end{equation}
when applied to $\hat{H}_{\rm{tot}}$ with $J_{1}=J_{2}=0$. We substitute the total Hamiltonian of Eq.~(\ref{eq:H_tot_long}) into the quantum channel $\mathcal{E}_{\text{CZ}}[\hat{\rho}]$, which we use in the optimization of the average gate fidelity and which we compare to the CZ gate in Eq.~(\ref{eq:CZ}), as displayed in Fig.~\ref{fig:CZ}(a).

As discussed before, we aim to minimize the gate time in order to limit the effect of dissipation on the gate performance. Unlike the iSWAP gate, the coupling constant does not depend on the qubit frequency. Both squeezing and the $1/\omega_m$ proportionality of the coupling constant benefit from low magnon frequencies, so that the gate time improves as the magnon frequency decreases as Fig.~\ref{fig:CZ}(a) shows. Thermal occupation at low frequencies, in turn, leads to a decrease of the average fidelity as for the iSWAP gate, see Fig.~\ref{fig:iSWAP}(a) for comparison, leading to a maximum of the average gate fidelity. As the magnon frequency is decreased, the parametric interaction term between magnons and qubits is resonantly enhanced (see Eq.~\eqref{eq:H_tot_long}), causing a small oscillating magnon occupation as seen in Fig.~\ref{fig:CZ}(b). The amplitude of these oscillations depends on the magnon frequency, resulting in the oscillations of the average fidelity as a function of frequency in  Fig.~\ref{fig:CZ}(a), with increasing amplitude for lower frequencies.

In order to obtain the results depicted in Fig.~\ref{fig:CZ} we set $a_J = 0$ and $\varphi_{b}=\pi/4$ for both qubits corresponding in this case to $\omega_{q}/(2\pi) = \SI{5.3}{GHz}$. We increased the Fock space size of the magnons to 6. We find an optimal average gate fidelity of $\bar{F}=99.43\%$ for $\omega_m /\omega_{q} = 0.027$. This corresponds to $g_{Z}/\omega_{q}=4.0\times10^{-4}$ ($g_{Z}/(2\pi) = \SI{2.1}{MHz}$), a thermal expectation number $n_{\text{th}} = 1$ and a squeezing enhancement of $e^{r} = 4.2$.

The time dynamics of the proposed gate is illustrated in Fig.~\ref{fig:CZ}(b) for the input state $\ket{1+} = \ket{1} \otimes  (\ket{0} + \ket{1})/\sqrt{2}$. For simplicity we take the first qubit to be the control qubit. Since this qubit is in the excited state, $\hat{\sigma}_z$ is applied to the target qubit according to Eq.~(\ref{eq:CZ}).
This gives $\hat{\sigma}_z \ket{+} = (\ket{0} - \ket{1})/\sqrt{2} = \ket{-}$. Due to $\hat{\sigma}_x \ket{+} = \ket{+}$ and $\hat{\sigma}_x \ket{-} = - \ket{-}$, we find that the expectation value of the second qubit changes from 1 at $t=0$ to -1 at $t=T_{Z}$, as Fig.~\ref{fig:CZ}(b) confirms.

The proposed gate of Eq.~(\ref{eq:gate}) assumes vacuum as the initial state for the magnons. Since the thermal expectation number is $n_{\text{th}} \sim 1$ for the optimal magnon frequency we found, a protocol for magnon cooling would need to be introduced in order to prepare the initial state into the ground state. To circumvent this necessity, we checked the performance of a gate using instead an initial magnon thermal state with $n_{\text{th}} = 0.99$.
Taking a magnon Fock space size of 12 in order to implement this thermal state in the simulations, we find an average gate fidelity $\bar{F} = 99.36 \%$.

We note that by flux-driving both qubits, a concept we introduce in the next section, we can effectively turn off the gate when required. In this case, the coupling constant $g_{Z}$ is proportional to $1/(\omega_{m} - \omega_{\text{ac}})$, where $\omega_{\text{ac}}$ is the driving frequency. Flux-driving at frequencies much larger than the magnon frequency can therefore be used to increase the detuning $\omega_{m} - \omega_{\text{ac}}$ such that the coupling is negligible, allowing to realize the gate efficiently by turning off the coupling in this manner after the gate time
\footnote{One could attempt to use flux-driving in order to increase the effective coupling strength, choosing $\omega_{\text{ac}}$ close to the magnon frequency. However, flux-driving also changes the coupling strengths $g_{1}$ and $g_{2}$, as will be shown in the next section. These new coupling strengths are typically lower than their non-driven equivalents. Ultimately, the fidelity values including driving turned out to be lower than without driving.}.

\subsection{iCNOT}

\begin{figure*}
\includegraphics[width=\linewidth]{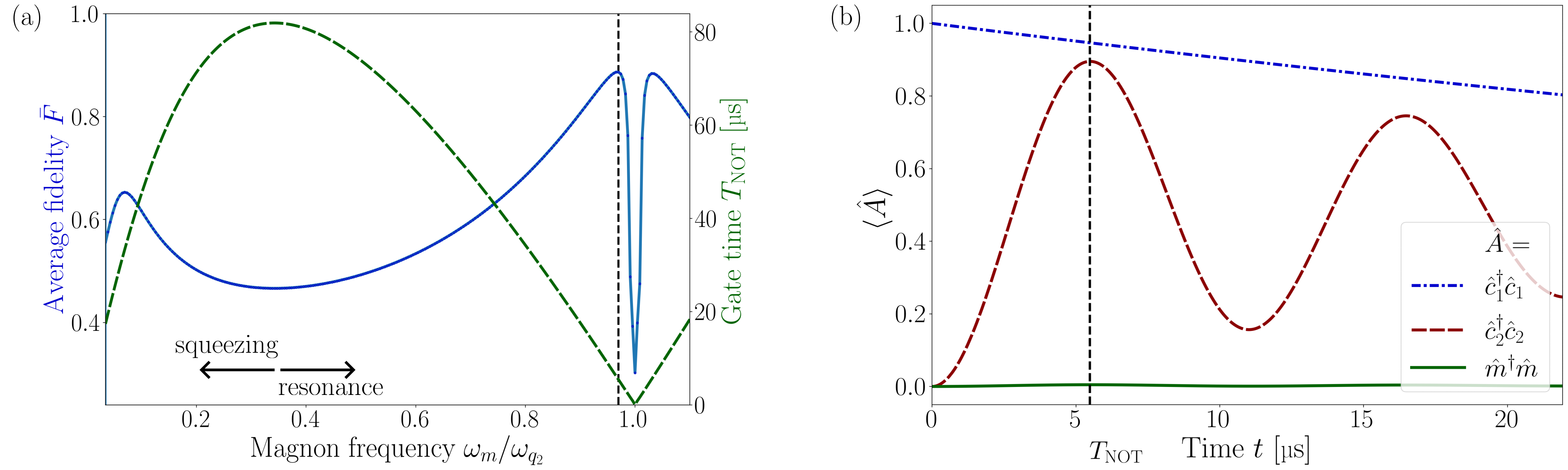}
\caption{\label{fig:iCNOT} (a) Average fidelity $\bar{F}$ (blue, solid) and gate time $T_{\rm{NOT}}$ (green, dashed) of the iCNOT gate as a function of the magnon frequency $\omega_{m}$.
The competition of magnon squeezing versus magnon-qubit resonance to optimize the gate fidelity is similar to the iSWAP gate of Fig.~\ref{fig:iSWAP}. The maximum fidelity, $\bar{F}=88.66\%$, is found at $\omega_m/\omega_{q_{2}} = 0.97$ (vertical dashed line).
(b) Qubit and magnon dynamics following the evolution with the Hamiltonian of Eq.~(\ref{eq:H_tot_opp}) including dissipation. We use the magnon frequency $\omega_m = 0.97 \, \omega_{q_{2}}$ found in (a). The occupations of both qubits (blue, dash-dotted and red, dashed) and magnons (green, solid) as a function of time are shown.
The target state is $\hat{U}_{\text{iCNOT}}\ket{10} = -i\ket{11}$ at the gate time $T_{\text{NOT}} = \SI{5.5}{\micro s}$ (vertical dashed line). Parameters are as in Fig~\ref{fig:iSWAP}.}
\end{figure*}

We consider the limit $a_{J_{1}}\rightarrow 0$ for one qubit, let it be qubit 1. For qubit 2 we set
$a_{J_{2}}\rightarrow 1$ and $\varphi_{b_{2}}=\pi/2$. This gives $J_{1}, g_{2} \rightarrow 0$. In order to make the interaction term leading to the iCNOT gate energetically allowed, in this case we need to consider a weak external ac bias  $\Phi_{b}=\Phi_{\text{ac}}\cos(\omega_{\text{ac}}t)$ with amplitude $\Phi_{\text{ac}}$ and frequency $\omega_{\text{ac}}$ applied to the first qubit. This changes the flux $\varphi_{b}\rightarrow\varphi_{\text{ac}}\cos(\omega_{\text{ac}}t)$, where $\varphi_{\text{ac}}=\pi\Phi_{\text{ac}}/\Phi_{0}$. For $\varphi_{\text{ac}}\ll1$ we find $g(\varphi_{b})\rightarrow \tilde{g}\cos(\omega_{\text{ac}}t)$,
with~\cite{Marios_setup} 
\begin{equation}
 \tilde{g}=-\frac{\mu_{0}I_{x}\mu_{\text{zpf}}e^{r}}{8\Phi_{0}d}\sqrt{8E_{C}E_{J}^{\Sigma}}\varphi_{\text{ac}} \,.  
\end{equation}

In the rotating frame of the drive we obtain
\begin{equation}
\begin{aligned}
  \hat{H}_{\text{RF}}^{\text{iCNOT}}
  & =\hbar\delta_{m}\hat{m}^{\dagger}\hat{m}+ \sum_{i=1,2}\hbar\delta_{q_{i}}\hat{c}_{i}^{\dagger}\hat{c}_{i} - \frac{E_{C}}{2} \hat{c}_{i}^{\dagger} \hat{c}_{i}^{\dagger}\hat{c}_{i} \hat{c}_{i}
  \\ & \quad \frac{\hbar\tilde{g}_{1}}{2}\hat{c}_{1}^{\dagger}\hat{c}_{1}
  (\hat{m}^{\dagger}+\hat{m}) + \hbar J_{2}(\hat{c}_{2}^{\dagger}\hat{m}+\hat{c}_{2}\hat{m}^{\dagger}),
\end{aligned}
\label{eq:H_RF_opp}
\end{equation}
where we defined $\delta_{m} = \omega_{m} - \omega_{\text{ac}}$ and $\delta_{q_{i}} = \omega_{q_{i}} - \omega_{\text{ac}}$ and used the rotating wave approximation, which is valid for $\tilde{g}_{1}\ll 4\omega_{\text{ac}}$. Performing a SW as detailed in Appendix~\ref{sec:SW_iCNOT} and choosing $\omega_{\text{ac}} = \omega_{q_{2}} + J_{2}^{2}/(\omega_{q_{2}} - \omega_m)$
yields
\begin{align}
    \hat{H}_{\text{SW}}^{\text{iCNOT}} & = \hbar\left(\delta_{m} - \frac{J_{2}^{2}}{\delta_{q_{2}} - \delta_{m}}\right)\hat{m}^{\dagger}\hat{m} \label{eq:H_SW_opp}
 \\ & \quad + \hbar\left(\delta_{q_{1}}-\frac{\tilde{g}_{1}^{2}}{4\delta_{m}}\right)\hat{\sigma}_{1}^{z} +\hbar\tilde{g}_{\text{NOT}}\hat{\sigma}_{1}^{z}(\hat{\sigma}_{2}^{+}+\hat{\sigma}_{2}^{-}) \nonumber
\end{align}
with the coupling strength
\begin{equation}
\tilde{g}_{\text{NOT}}=\frac{\tilde{g}_{1} J_{2}}{4}\left(\frac{1}{\delta_{q_{2}}-\delta_{m}}-\frac{1}{\delta_{m}}\right).
\end{equation}
The frequency of the flux drive $\omega_{\text{ac}}$ is chosen by matching the Stark-shifted frequency of the second qubit in order to make the interaction term resonant. 
The SW transformation is valid for $J_{2}\ll\omega_{q_{2}}-\omega_{m}$ and $\tilde{g}_{1}\ll 2\delta_{m}$. Time propagation of the coupling term of the effective Hamiltonian of Eq.~(\ref{eq:H_eff_opp}) up to $T_{\text{NOT}}=\pi/(2|\tilde{g}_{\text{NOT}}|)$ gives an iCNOT gate
\begin{equation}
  \hat{U}_{\text{iCNOT}}=\ket{0}\bra{0} \otimes \hat{I} \mp i \ket{1}\bra{1} \otimes\hat{\sigma}_{x},
\label{eq:iCNOT}
\end{equation}
since $\exp\bigl(-i\tilde{g}_{\text{NOT}}T_{\text{NOT}}\hat{\sigma}_{1}^{z}(\hat{\sigma}_{2}^{+}+\hat{\sigma}_{2}^{-})\bigl) = \hat{U}_{\text{iCNOT}}$.
This gate resembles a CNOT gate, but adds a phase $\mp i$ if the control qubit, i.e. the first qubit, is excited. Thus, we denominate it an iCNOT gate. The sign of the phase $\mp i$ corresponds to $\tilde{g}_{\text{NOT}} \gtrless 0$.

Similarly as performed for the previous gates, in order to compensate for single qubit rotations we cancel the diagonal term of the control qubit of Eq.~(\ref{eq:H_SW_opp}) by performing the unitary rotation 
\begin{equation}
    \hat{U}_{\text{SQ}}(t)=\exp\biggl[it\biggl(\delta_{q_{1}}-\frac{\tilde{g}_{1}^{2}}{4\delta_{m}}\biggr)\hat{\sigma}_{1}^{z}\biggr].
\end{equation}
For Eq.~(\ref{eq:H_SW_opp}) we find
\begin{equation}  
\hat{H}_{\text{eff}}^{\text{iCNOT}}=\hbar\left(\delta_{m} - \frac{J_{2}^{2}}{\delta_{q_{2}} - \delta_{m}}\right)\hat{m}^{\dagger}\hat{m} + \hbar\tilde{g}_{\text{NOT}}\hat{\sigma}_{1}^{z}(\hat{\sigma}_{2}^{+} + \hat{\sigma}_{2}^{-}).
\label{eq:H_eff_opp}
\end{equation}
Rotating the Hamiltonian before the SW transformation of Eq.~(\ref{eq:H_RF_opp}) with the same unitary transformation yields 
\begin{align}
    \hat{H}_{\text{tot}}^{\text{iCNOT}} & = \hbar\delta_{m}\hat{m}^{\dagger}\hat{m}+\frac{\hbar\tilde{g}_{1}^{2}}{4\delta_{m}}\hat{c}_{1}^{\dagger}\hat{c}_{1} + \hbar\delta_{q_{2}}\hat{c}_{2}^{\dagger}\hat{c}_{2} \nonumber \\ & \quad \label{eq:H_tot_opp}
    -\sum_{i=1,2} \frac{E_{C}}{2} \hat{c}_{i}^{\dagger}\hat{c}_{i}^{\dagger}\hat{c}_{i}\hat{c}_{i}  
    \\ & \quad +\frac{\hbar\tilde{g}_{1}}{2}\hat{c}_{1}^{\dagger}\hat{c}_{1}(\hat{m}^{\dagger}+\hat{m})+\hbar J_{2}(\hat{c}_{2}^{\dagger}\hat{m}+\hat{c}_{2}\hat{m}^{\dagger}). \nonumber
\end{align}

For the channel $\mathcal{E}_{\text{iCNOT}}[\hat{\rho}]$ we use the total Hamiltonian of Eq.~(\ref{eq:H_tot_opp}). We compute the average fidelity of the proposed gate $\mathcal{E}_{\text{iCNOT}}[\hat{\rho}]$ with the ideal gate $\hat{U}_{\text{iCNOT}}$, which can be found in Eq~(\ref{eq:iCNOT}). The result is shown in Fig.~\ref{fig:iCNOT}(a). The dependence of the average gate fidelity on the magnon frequency is akin to the iSWAP gate. The coupling constant $\tilde{g}_{\text{NOT}}$ reminds us of a combination of $g_{S}$ and $g_{Z}$ of Eqs.~(\ref{eq:g_xy}) and (\ref{eq:g_zz}), respectively.
Since $\delta_{q_{2}} - \delta_{m} = \omega_{q_{2}} - \omega_{m}$, magnon frequencies close to the frequency of the target qubit give rise to large coupling strengths, but approach the breakdown of the SW transformation, as signalled by a vanishing gate time in Fig.~\ref{fig:iCNOT}(a).
Also, the factor $\delta_{m}$ is small for these magnon frequencies due to $\omega_{\text{ac}} \approx \omega_{q_{2}}$.
Similarly to the iSWAP gate, magnon frequencies in this regime lead to little increase of the coupling constant due to squeezing: $\tilde{g}_{\text{NOT}}\propto e^{2r} \approx \SI{1}{}$. In order to increase the coupling constant through squeezing one needs $\omega_m \ll \gamma\mu_{0}M_{s}$. Thus, the competition to maximize the coupling constant and hence to restrict dissipative processes is similar to the iSWAP gate.

For the results displayed in Fig.~\ref{fig:iCNOT} we set the asymmetry parameter and reduced flux of the first (second) qubit to $a_{J_{1}} = 0$ and $\varphi_{\text{ac}_{1}}=\pi/10$ ($a_{J_{2}} = 0.9$ and $\varphi_{b_{2}} = \pi/2$), corresponding to $\omega_{q_{1}} = \SI{6.2}{GHz}$ ($\omega_{q_{2}} = \SI{6.0}{GHz}$). We find a maximum average fidelity of $\bar{F}=88.66\%$ for $\omega_m/\omega_{q_{2}} = 0.97$, at an effective coupling strength of  $\tilde{g}_{\text{NOT}}/\omega_{q_{2}} = 7.6\times10^{-6}$ corresponding to $\tilde{g}_{\text{NOT}}/(2\pi) = \SI{46}{kHz}$.

Fig.~\ref{fig:iCNOT}(b) displays the dynamics of the system for the magnon frequency which maximizes the average fidelity and $\ket{10}$ as our input state. At $t=T_{\text{NOT}}$ we should find the state $\hat{U}_{\text{iCNOT}}\ket{10} = -i\ket{11}$.
However, due to a relatively large gate time $T_{\text{NOT}} = \SI{5.5}{\micro s}$ compared to decay $T_1$ and dephasing $T_{\phi}$, dissipation has a relatively large influence. Therefore, we see that the control qubit, i.e. the first qubit, has lost some of its initial excitation. The target qubit does not reach $\langle \hat{c}^{\dagger}_{2} \hat{c}_{2} \rangle = 1$ either. Turning off the coupling after the gate time $T_{\text{NOT}}$ can be simply achieved by switching off the ac driving of the control qubit.

Although the iSWAP and iCNOT gate have some resemblances regarding the effective coupling strength $\tilde{g}_{\text{NOT}}$, the iCNOT gate does not achieve a similar fidelity. The reason is that $\tilde{g}_{\text{NOT}}$ is proportional to $\tilde{g}_{1}$, which is a factor 10 smaller than its iSWAP equivalent $J_{1}$, leading to a more detrimental effect of dissipation.

\section{Conclusions}\label{sec:con}

We have theoretically demonstrated that magnons can be used to mediate strong qubit-qubit coupling, where for feasible experimental parameters we obtained coupling strengths surpassing the qubit dissipation.
The coherent magnon-qubit exchange interaction and radiation-pressure interaction can be adopted to engineer two-qubit quantum gates. With the exchange interaction an iSWAP gate is realized by using highly asymmetric SQUIDs. The non-linear interaction generated by symmetric SQUIDs realizes a CZ gate. By combining the exchange interaction on a highly asymmetric SQUID and the radiation-pressure on a symmetric SQUID an iCNOT gate is implemented. We numerically tested these proposed gates under realistic experimental conditions and find an average gate fidelity which equals $99.00 \%$ for the iSWAP gate, $99.43\%$ for the CZ gate and $88.66 \%$ for the iCNOT gate.
The coupling strenghts with respect to the qubit dissipation equal $T_{1} g_{S} / (2\pi) = 49$ for the iSWAP, $T_{1} g_{Z} / (2\pi) = 214$ for the CZ and $T_{1} \tilde{g}_{\text{NOT}} / (2\pi) = 4.6$ for the iCNOT gate.
Furthermore, we found no leakage out of the computational space for both qubits.

In all of our simulations, we assumed the initial magnonic state to be vacuum instead of a thermal state for computational simplicity. 
We found that the optimal magnon frequency for the iSWAP and iCNOT gate is in the GigaHertz regime, and thus vacuum state assumption can be achieved passively by cooling the system down to temperatures $< \SI{50}{\milli\kelvin}$.
However, the optimal magnon frequency we found for the CZ gate is in the MegaHertz regime, leading to a sizeable thermal population of magnons for the temperature used in the simulations ($\SI{10}{\milli\kelvin}$).
In order to verify the validity of our results, we simulated the CZ gate further with an initial magnon thermal state dictated by the optimal magnon frequency at the given temperature, and found a similar average fidelity $\bar{F} = 99.36 \%$.

While the average gate fidelities for the iSWAP and CZ gates are above the error correction threshold~\cite{Fowler2012F_thresh2, Fowler2012F_thresh3}, this is not the case for the iCNOT, for which the gate time is not small with respect to the qubit relaxation timescales.
One could improve this by introducing waveguides that transport the electromagnetic wave emitted by the magnet to the SQUID loop~\cite{Yu2020chirality1}, thereby increasing the total flux through the loop and, as a result, the coupling.
Moreover, we have shown that the qubit-magnon coupling strength can be enhanced with magnetization squeezing for magnon frequencies much lower than $\mu_0\gamma_0 M_s$, and using qubit frequencies close to this regime.
However, transmons typically operate at higher frequencies in the 4-8~GHz regime~\cite{sqrtSWAP}.
Therefore, the performance of the iSWAP and iCNOT gates could be further improved in qubit-magnon hybrid systems involving low-frequency qubits~\cite{vool2017driving,gely2019observation}. For lower operating frequencies the impact of thermal occupation should be evaluated.

\begin{acknowledgments}
S.S., M.K., and S.V.K. acknowledge financial support by the German Federal Ministry of Education and Research (BMBF) project QECHQS (Grant No. 16KIS1590K). S.V.K. and Y.M.B. acknowledge financial support by the EU-Project No. HORIZON-EIC-2021-PATHFINDEROPEN- 01 PALANTIRI-101046630.
\end{acknowledgments}

\FloatBarrier

\appendix

\begin{widetext}

\section{Geometrical factor}\label{app:geo}

With the scalar potential $P(\mathbf{r})$ caused by the magnetization and which obeys $\mathbf{B} = - \mu_{0} \nabla P(\mathbf{r})$, we describe the magnetic stray field $\mathbf{B}$.
We decompose the potential in terms which are given rise to by the magnetic moment of each Cartesian component, such that
\begin{equation}
    P(\mathbf{r}) = \sum_{i=x,y,z} \mu_{i} p_{i}(\mathbf{r}).
\end{equation}
We use the magnetic field which is caused by $\mu_{z}$ to obtain Fig.~\ref{fig:B_crit}, where we plotted the $z$ component of this field and used $\mu_{z} = M_{s} V_{m}$.
We set the magnetic moment of the $x$ and $y$ component equal to the magnetic moment fluctuations of Eq.~(\ref{eq:mu_x}) and treat them as operators, so $\mu_{x} = \Delta \hat{\mu}_{x}$ and $\mu_{y} = \Delta \hat{\mu}_{y}$. We determine the flux through the SQUID loop caused by these fluctuations with
\begin{equation}
    \Phi (\Delta \hat{\boldsymbol{\mu}}) = -\mu_{0}\sum_{i=x,y}\Delta\hat{\mu}_{i}\int_{\mathrm{SQUID}} \nabla p_{i} (\mathbf{r})  \cdot \text{d}\boldsymbol{A}.
\end{equation}
By comparing this relation with Eq.~(\ref{eq:phasecoupling}), we find
\begin{equation}
    I_{i}=-4\pi\int_{\mathrm{SQUID}} \nabla p_{i} (\mathbf{r})  \cdot \text{d}\boldsymbol{A}.
\end{equation}
Due to the symmetry of the setup we find $I_{y} = 0$.

As described in the main text, a limiting factor of $I_{x}$ is the superconducting critical field $B_{c}$. The amplitude of the stray magnetic field in the $z$ direction, $B_{z}$, at $z=d$ should be smaller than $B_{c}$. For the magnet with an ellipsoidal shape, the field $B_{z}(z=L_{z})$ is about two orders of magnitude lower than the critical field of typical superconductors~\cite{Popinciuc2012Criticalfield}, as one can see in Fig.~\ref{fig:B_crit}. Thus, the SQUID loop can be positioned at touching distance from the magnet and hence we set $d - L_{z} = \SI{10}{nm}$. Furthermore, $L_{x}$ and $L_{z}$ should be chosen such that $N_{T}\approx 1/2$. However, $L_{x} \gg L_{z}$ resembles a magnet infinitely stretched along $x$ axis, which is known to have less magnetic field in its direct vicinity in the $x=0$ plane than a spherical magnet. Therefore, we set $N_{T}=0.45$, which fixes the ratio of $L_{x}$ and $L_{z}$. Increasing the loop radius $R$ leads to an increase in the coupling strength, yet gives rise to more qubit noise. We put $R=\SI{25}{\micro m}$. By varying $L_{x}$ we find an optimal effective coupling constant for $L_{x} = \SI{16}{\micro m}$ and $L_{z} = \SI{3.9}{\micro m}$. This corresponds to $I_{x} = -\SI{0.12}{}/\SI{}{\micro m}$. Note that coherent magnon quantum states have been demonstrated in magnets of sizes up to $\SI{1}{mm}$~\cite{Xu2023transphotmag4}, significantly larger than the sizes considered here.

Considering a sphere with the same volume and with radius $\tilde{r}$ such that $\tilde{r}^{3}= L_{x} L_{z}^{2}$, while fixing $d - \tilde{r} = \SI{10}{nm}$, gives an effective coupling constant which is a factor $2.8$ higher than the coupling constant for an ellipsoidal magnet excluding the squeezing enhancement.

\begin{figure}
\includegraphics[width=\linewidth]{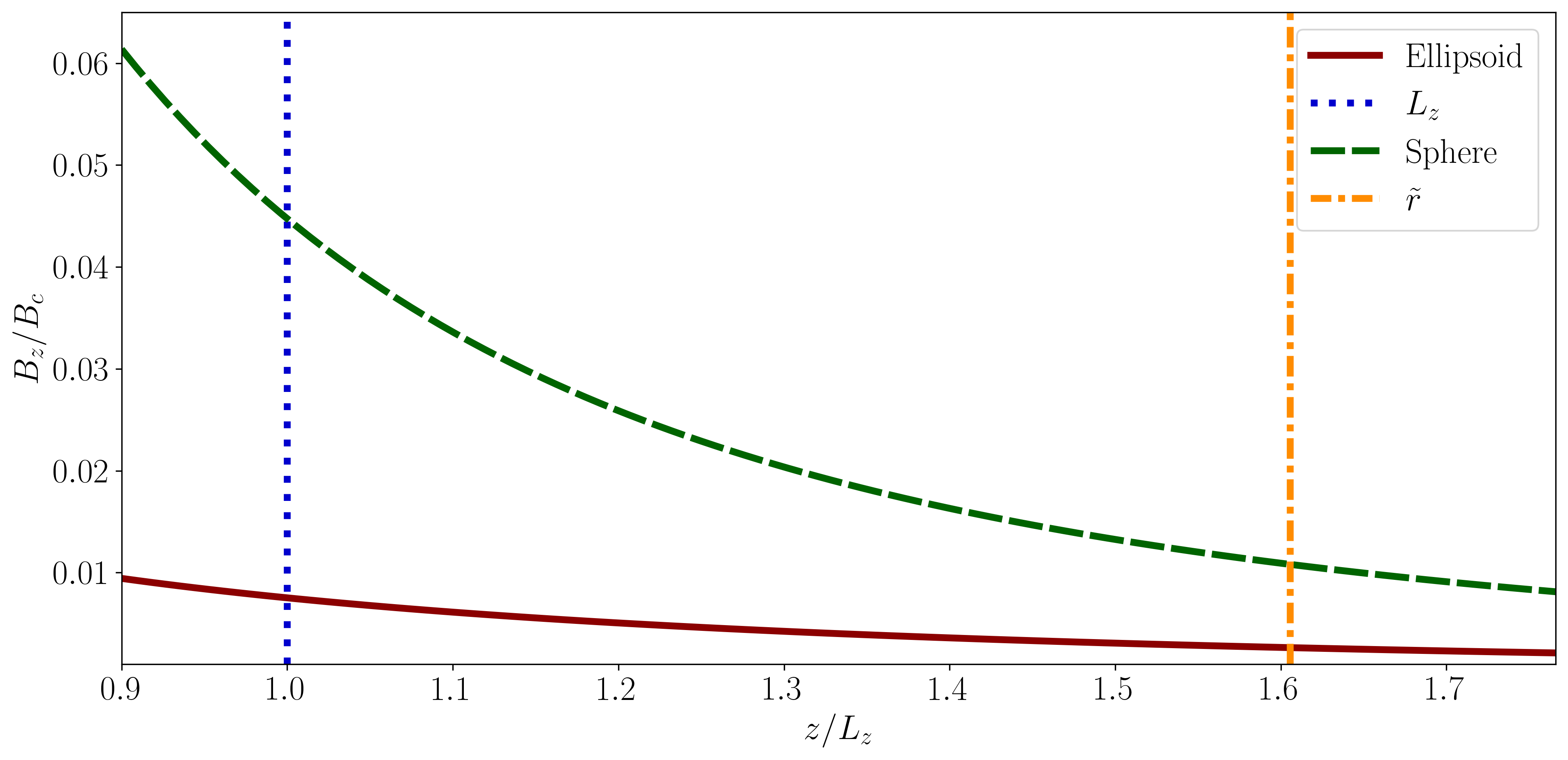}
\caption{\label{fig:B_crit} Magnetic field in $z$ direction in units of the critical field $B_{c}$ for an ellipsoidal and spherical magnet as a function of the distance $z$ in units of $L_{z}$.}
\end{figure}

\section{Average gate fidelity}\label{app:fid}

To determine the average gate fidelity efficiently we use the following relation \cite{fidelity},
\begin{equation}
  \bar{F}\bigl(\mathcal{E},\hat{U}\bigl)=\frac{\sum_{j}\text{tr}\left[\hat{U}_{j}^{\dagger}\hat{U}^{\dagger}\mathcal{E}[\hat{U}_{j}]\hat{U}\right]+d^{2}}{d^{2}(d+1)},
\end{equation}
where the unitary operators $\hat{U}_{j}$ form an orthogonal basis and $d=4$ is the dimension of the two-qubit space. We choose $\hat{U}_{j}=\hat{\sigma}_{k}\otimes\hat{\sigma}_{l}$, where $\hat{\sigma}_{m}$ is identity or a Pauli matrix, so $\hat{\sigma}_{m} \in \{\hat{I},\hat{\sigma}_x,\hat{\sigma}_y, \hat{\sigma}_z \}$.
The input of $\mathcal{E} [ \hat{\rho} ]$ is restricted to density matrices. Since the Pauli matrices have zero trace, we write these in terms of the density matrices.
We find $\hat{I} = \hat{\rho}_0 + \hat{\rho}_1$, $\hat{\sigma}_x = 2\hat{\rho}_{+}-\hat{\rho}_{0}-\hat{\rho}_{1}$, $\hat{\sigma}_y = 2\hat{\rho}_{i-}-\hat{\rho}_{0}-\hat{\rho}_{1}$ and $\hat{\sigma}_z = \hat{\rho}_0 - \hat{\rho}_1$. Here, $\hat{\rho}_i = \ket{i}\bra{i}$, $\ket{+}=\frac{1}{\sqrt{2}}(\ket{0} + \ket{1} )$ and $\ket{i-}=\frac{1}{\sqrt{2}}(\ket{0} - i\ket{1} )$.
We use the linearity of $\mathcal{E}[\hat{\rho}]$ to rewrite the expressions. For example, one finds
\begin{equation}
  \mathcal{E} [\hat{I} \otimes \hat{\sigma}_x ] = 2\mathcal{E}[\hat{\rho}_{0}\otimes\hat{\rho}_{+}]-\mathcal{E}[\hat{\rho}_{0}\otimes\hat{\rho_{0}}]-\mathcal{E}[\hat{\rho_{0}}\otimes\hat{\rho}_{1}]+2\mathcal{E}[\hat{\rho}_{1}\otimes\hat{\rho}_{+}]-\mathcal{E}[\hat{\rho}_{1}\otimes\hat{\rho}_{0}]-\mathcal{E}[\hat{\rho}_{1}\otimes\hat{\rho}_{1}].  
\end{equation}

\section{Schrieffer-Wolff transformation}\label{app:SW}
    
With a SW transformation one transforms a Hamiltonian $\hat{H}$ according to $\hat{H}_{\text{SW}}=e^{-\hat{S}}\hat{H}e^{\hat{S}}$
with an anti-Hermitian generator $\hat{S}=-\hat{S}^{\dagger}$. We write
$\hat{H}=\hat{H}_{0}+\hat{H}_{\text{int}}$, where $\hat{H}_{0}$ is already
diagonalized with respect to the tensor product of the number basis for qubits and magnon, on the contrary to $\hat{H}_{\text{int}}$. We assume that both $\hat{S}$ and $\hat{H}_{\text{int}}$ are proportional
to a coupling constant (e.g. $J$ or $g$). We approximate to second order in this constant. Using the Baker-Campbell-Hausdorff
formula to second order gives
\[
\hat{H}_{\text{SW}} \approx\hat{H}_{0}+\hat{H}_{\text{int}}+\left[\hat{S},\hat{H}_{0}\right]+\left[\hat{S},\hat{H}_{\text{int}}\right]+\frac{1}{2}\left[\hat{S},\left[\hat{S},\hat{H}_{0}\right]\right].
\]
By imposing $\left[\hat{S},\hat{H}_{0}\right]=-\hat{H}_{\text{int}}$
one gets
\[
\hat{H}_{\text{\text{SW}}}=\hat{H}_{0}+\frac{1}{2}\left[\hat{S},\hat{H}_{\text{int}}\right].
\]
This approximation is valid for $|\hat{S}|\ll1$.

The diagonalized Hamiltonian can be found in Eq.~(\ref{eq:H_0}) and the interaction Hamiltonian in Eq.~(\ref{eq:H_int}). We use the generator $\hat{S} = \hat{S}_{J} + \hat{S}_{g}$, where
\begin{equation}
    \hat{S}_{J} = \sum_{i=1,2}J_{i}(\chi_{i}(\hat{c}_{i}^{\dagger}\hat{c}_{i})\hat{c}_{i}^{\dagger}\hat{m}-\hat{c}_{i}\hat{m}^{\dagger}\chi_{i}(\hat{c}_{i}^{\dagger}\hat{c}_{i})),
\end{equation}
with transmon susceptibility
\begin{equation}
    \chi_{i}(\hat{c}_{i}^{\dagger}\hat{c}_{i})=\frac{1}{\omega_{m}-\omega_{q_{i}}+\frac{E_{C_{i}}}{\hbar}(\hat{c}_{i}^{\dagger}\hat{c}_{i}-1)},
\end{equation}
and
\begin{equation}
    \hat{S}_{g} = \sum_{i=1,2}\frac{g_{i}}{\omega_{m}}\hat{c}_{i}^{\dagger}\hat{c}_{i}(\hat{m}^{\dagger}-\hat{m}).
\end{equation}
We compute
\begin{equation}
\begin{aligned}
\left[\hat{S}_{J},\hat{H}_{0}\right]
  & = \sum_{i=1,2}\biggl[J_{i}(\chi_{i}(\hat{c}_{i}^{\dagger}\hat{c}_{i})\hat{c}_{i}^{\dagger}\hat{m}-\hat{c}_{i}\hat{m}^{\dagger}\chi_{i}(\hat{c}_{i}^{\dagger}\hat{c}_{i})),\hbar\omega_{q_{i}}\hat{c}_{i}^{\dagger}\hat{c}_{i} -\frac{E_{C}}{2} \hat{c}_{i}^{\dagger}\hat{c}_{i}^\dagger\hat{c}_{i}\hat{c}_{i} \biggl]
 \\ & \quad+\sum_{i=1,2}\left[J_{i}(\chi_{i}(\hat{c}_{i}^{\dagger}\hat{c}_{i})\hat{c}_{i}^{\dagger}\hat{m}-\hat{c}_{i}\hat{m}^{\dagger}\chi_{i}(\hat{c}_{i}^{\dagger}\hat{c}_{i})),\hbar\omega_{m}\hat{m}^{\dagger}\hat{m}\right]\\
 & =-\sum_{i=1,2}\hbar J_{i}(\hat{c}_{i}\hat{m}^{\dagger}+\hat{c}_{i}^{\dagger}\hat{m}) = - \sum_{i=1,2} \hat{H}_{J}^{i}
\end{aligned}
\end{equation}
and
\begin{equation}
\left[\hat{S}_{g},\hat{H}_{0}\right] =\sum_{i=1,2}\hbar g_{i}\hat{c}_{i}^{\dagger}\hat{c}_{i}[\hat{m}^{\dagger}-\hat{m},\hat{m}^{\dagger}\hat{m}] =-\sum_{i=1,2}\hbar g_{i}\hat{c}_{i}^{\dagger}\hat{c}_{i}(\hat{m}^{\dagger}+\hat{m}) = - \sum_{i=1,2} \hat{H}_{g}^{i}.
\end{equation}
Thus, we verify
\begin{equation}
    \left[\hat{S},\hat{H}_{0}\right] =\left[\hat{S}_{J},\hat{H}_{0}\right]+\left[\hat{S}_{g},\hat{H}_{0}\right] = - \sum_{i=1,2} \hat{H}_{J}^{i} + \hat{H}_{g}^{i} =-\hat{H}_{\text{int}}.
\end{equation}
This leaves us to determine
\begin{equation}
    \left[\hat{S},\hat{H}_{\text{int}}\right] = \hat{H}_{J,J} + \hat{H}_{J,g} + \hat{H}_{g,J} + \hat{H}_{g,g},
\end{equation}
with $\hat{H}_{J,J} =\left[\hat{S}_{J},\sum_{i=1,2}\hat{H}_{J}^{i}\right]$, $\hat{H}_{J,g} = \left[\hat{S}_{J},\sum_{i=1,2}\hat{H}_{g}^{i}\right]$, $\hat{H}_{g,J} =\left[\hat{S}_{g},\sum_{i=1,2}\hat{H}_{J}^{i}\right]$ and $\hat{H}_{g,g} = \left[\hat{S}_{g},\sum_{i=1,2}\hat{H}_{g}^{i}\right]$.
We find
\begin{equation}
\begin{aligned}
\hat{H}_{J,J}
 & = \sum_{i=1,2}\hbar J_{i}^{2}\biggl( 2 \chi_{i}(\hat{c}_{i}^{\dagger}\hat{c}_{i})\hat{c}_{i}^{\dagger}\hat{c}_{i} -\left(2\chi_{i}(\hat{c}_{i}^{\dagger}\hat{c}_{i}+1)(1-\hat{c}_{i}^{\dagger}\hat{c}_{i})+2\chi_{i}(\hat{c}_{i}^{\dagger}\hat{c}_{i})\hat{c}_{i}^{\dagger}\hat{c}_{i}\right)\hat{m}^{\dagger}\hat{m} \\ & \quad \quad + \left(\chi_{i}(\hat{c}_{i}^{\dagger}\hat{c}_{i})-\chi_{i}(\hat{c}_{i}^{\dagger}\hat{c}_{i}-1)\right)(\hat{c}_{i}^{\dagger})^{2}\hat{m}^{2} +\left(\chi_{i}(\hat{c}_{i}^{\dagger}\hat{c}_{i}+2) -\chi_{i}(\hat{c}_{i}^{\dagger}\hat{c}_{i}+1) \right )\hat{c}_{i}^{2}(\hat{m}^{\dagger})^{2} \biggr)
 \\ & \quad +\hbar J_{1}J_{2}\left(\chi_{1}(\hat{c}_{1}^{\dagger}\hat{c}_{1})\hat{c}_{1}^{\dagger}\hat{c}_{2}+\hat{c}_{1}\chi_{1}(\hat{c}_{1}^{\dagger}\hat{c}_{1})\hat{c}_{2}^{\dagger}\right) +\hbar J_{1}J_{2}\left(\chi_{2}(\hat{c}_{2}^{\dagger}\hat{c}_{2})\hat{c}_{2}^{\dagger}\hat{c}_{1}+\hat{c}_{2}\chi_{2}(\hat{c}_{2}^{\dagger}\hat{c}_{2})\hat{c}_{1}^{\dagger}\right),
\end{aligned}
\label{eq:H_JJ}
\end{equation}
\begin{equation}
\begin{aligned}
    \hat{H}_{J,g} & =\sum_{i=1,2}\hbar J_{i}g_{i}\left(\chi_{i}(\hat{c}_{i}^{\dagger}\hat{c}_{i})\hat{c}_{i}^{\dagger}(\hat{c}_{i}^{\dagger}\hat{c}_{i}-(\hat{m}+\hat{m}^{\dagger})\hat{m}) +\chi_{i}(\hat{c}_{i}^{\dagger}\hat{c}_{i}+1)\hat{c}_{i}(\hat{c}_{i}^{\dagger}\hat{c}_{i}-(\hat{m}+\hat{m}^{\dagger})\hat{m}^{\dagger})\right) \\ & \quad +\hbar J_{1}g_{2}\hat{c}_{2}^{\dagger}\hat{c}_{2}\left(\chi_{1}(\hat{c}_{1}^{\dagger}\hat{c}_{1})\hat{c}_{1}^{\dagger}+\hat{c}_{1}\chi_{1}(\hat{c}_{1}^{\dagger}\hat{c}_{1})\right) + \hbar g_{1}J_{2}\hat{c}_{1}^{\dagger}\hat{c}_{1}\left(\chi_{2}(\hat{c}_{2}^{\dagger}\hat{c}_{2})\hat{c}_{2}^{\dagger}+\hat{c}_{2}\chi_{2}(\hat{c}_{2}^{\dagger}\hat{c}_{2})\right),
\end{aligned}
\label{eq:H_Jg}
\end{equation}
\begin{equation}
\begin{aligned}
    \hat{H}_{g,J} & =\sum_{i=1,2}\frac{\hbar J_{i}g_{i}}{\omega_{m}}\Big((\hat{c}_{i}^{\dagger}+\hat{c}_{i})(\hat{m}^{\dagger}\hat{m}+1) -\hat{c}_{i}(\hat{m}^{\dagger})^{2}-\hat{c}_{i}^{\dagger}\hat{m}^{2} -\hat{c}_{i}^{\dagger}\hat{c}_{i}\hat{c}_{i}^{\dagger}-\hat{c}_{i}\hat{c}_{i}^{\dagger}\hat{c}_{i}\Big) \\ & \quad -\frac{\hbar g_{1}J_{2}}{\omega_{m}}\hat{c}_{1}^{\dagger}\hat{c}_{1}(\hat{c}_{2}^{\dagger}+\hat{c}_{2})-\frac{\hbar J_{1}g_{2}}{\omega_{m}}(\hat{c}_{1}^{\dagger}+\hat{c}_{1})\hat{c}_{2}^{\dagger}\hat{c}_{2}
\end{aligned}
\label{eq:H_gJ}
\end{equation}
and finally
\begin{equation}
 \hat{H}_{g,g}
 =-\sum_{i=1,2}\frac{2\hbar g_{i}^{2}}{\omega_{m}}(\hat{c}_{i}^{\dagger}\hat{c}_{i})^{2}-\frac{4\hbar g_{1}g_{2}}{\omega_{m}}\hat{c}_{1}^{\dagger}\hat{c}_{1}\hat{c}_{2}^{\dagger}\hat{c}_{2}.
\label{eq:H_gg}
\end{equation}
Thus, the transformed Hamiltonian we obtain is
\begin{equation}
    \hat{H}_{\text{SW}} = \hat{H}_{0} + \frac{1}{2}\left(\hat{H}_{J,J} + \hat{H}_{J,g} + \hat{H}_{g,J} + \hat{H}_{g,g}\right).
\label{eq:H_SW}
\end{equation}
We note that Eq.~(\ref{eq:H_JJ}) contains a SWAP-like qubit-qubit term, Eqs.~(\ref{eq:H_Jg}) and~(\ref{eq:H_gJ}) show a controlled-SWAP form and Eq.~(\ref{eq:H_gg}) a controlled-phase term. By choosing different combinations of coupling constants $J$ and $g$, we can tune between these interactions.

\subsection{iSWAP}\label{sec:SW_iSWAP}
For $g_{1}=g_{2}=0$ we find $\hat{H}_{J,g} = \hat{H}_{g,J} = \hat{H}_{g,g} = 0$. Therefore, we find
\begin{equation}
    \hat{H}_{\text{SW}}^{\text{iSWAP}} = \hat{H}_{0} + \frac{1}{2}\hat{H}_{J,J}.
\end{equation}
Truncating the higher qubit levels, such that only the ground and first excited states remain, gives
\begin{equation}
    \hat{H}_{J,J} = \sum_{i=1,2} 2 \hbar J_{i}^{2} \chi_{i}(1)(\hat{\sigma}_{i}^{z}-\hat{m}^{\dagger}\hat{m}) + \hbar J_{1}J_{2}(\chi_{1}(1)+\chi_{2}(1))(\hat{\sigma}_{1}^{+}\hat{\sigma}_{2}^{-}+\hat{\sigma}_{1}^{-}\hat{\sigma}_{2}^{+}).
\end{equation}
This leads to Eq.~(\ref{eq:H_SW_trans}).

\subsection{CZ}\label{sec:SW_CZ}
For $J_{1}=J_{2}=0$ we have $\hat{H}_{J,J} = \hat{H}_{J,g} = \hat{H}_{g,J} = 0$. This gives
\begin{equation}
    \hat{H}_{\text{SW}}^{\text{iSWAP}} = \hat{H}_{0} + \frac{1}{2}\hat{H}_{g,g}.
\end{equation}
Taking only the ground and first state of the qubit gives 
\begin{equation}
    \hat{H}_{g,g} =-\sum_{i=1,2}\frac{2\hbar g_{i}^{2}}{\omega_{m}}\hat{\sigma}_{i}^{z}-\frac{4\hbar g_{1}g_{2}}{\omega_{m}}\hat{\sigma}_{1}^{z}\hat{\sigma}_{2}^{z},
\end{equation}
since $(\hat{\sigma}_{i}^{z})^2 = \hat{\sigma}_{i}^{z}$.
This yields Eq.~(\ref{eq:H_SW_long}).

\subsection{iCNOT}\label{sec:SW_iCNOT}

We set $J_{1}=g_{2} = 0$. Since $\hat{H}_{J,J}$, $\hat{H}_{J,g}$, $\hat{H}_{g,J}$ and $\hat{H}_{g,g}$ do not vanish, the transformed Hamiltonian is given by Eq.~(\ref{eq:H_SW}).
From now on, we limit the qubit to its lowest two levels.
Eq.~(\ref{eq:H_RF_opp}) shows that we can use the results of the SW transformation if we implement following substitutions: $\omega_{m} \rightarrow \delta_{m}$, $\omega_{q_{i}} \rightarrow \delta_{q_{i}}$ and $g_{1} \rightarrow \tilde{g}_{1}/2$. We find
\begin{equation}
    \hat{H}_{J,J} = \frac{ 2 \hbar J_{2}^{2} }{\delta_{m} - \delta_{q_{i}}} (\hat{\sigma}_{2}^{z}-\hat{m}^{\dagger}\hat{m}),
\end{equation}
\begin{equation}
    \hat{H}_{J,g} = \frac{\hbar \tilde{g}_{1}J_{2}}{2(\delta_{q_{2}}-\delta_{m})}\hat{\sigma}_{1}^{z}(\hat{\sigma}_{2}^{+}+\hat{\sigma}_{2}^{-}), 
\end{equation}
\begin{equation}
   \hat{H}_{g,J} = - \frac{\hbar \tilde{g}_{1}J_{2}}{2\delta_{m}}\hat{\sigma}_{1}^{z}(\hat{\sigma}_{2}^{+}+\hat{\sigma}_{2}^{-})
\end{equation}
and
\begin{equation}
    \hat{H}_{g,g} = -\frac{\hbar \tilde{g}_{1}^{2}}{2\delta_m} \hat{\sigma}_{1}^{z}.
\end{equation}
With Eq.~(\ref{eq:H_SW}) we find
\begin{equation}
\begin{aligned}
  \hat{H}_{\text{SW}}^{\text{iCNOT}} & = \hbar\left(\delta_{m}-\frac{J_{2}^{2}}{\delta_{q_{2}}-\delta_{m}}\right)\hat{m}^{\dagger}\hat{m} +\hbar\left(\delta_{q_{1}}-\frac{\tilde{g}_{1}^{2}}{4\delta_{m}}\right)\hat{\sigma}_{1}^{z} +\hbar\left(\delta_{q_{2}}+\frac{J_{2}^{2}}{\delta_{q_{2}}-\delta_{m}}\right)\hat{\sigma}_{2}^{z} \\
  & \quad + \frac{ \hbar \tilde{g}_{1} J_{2}}{4}\left(\frac{1}{\delta_{q_{2}}-\delta_{m}}-\frac{1}{\delta_{m}}\right) \hat{\sigma}_{1}^{z} (\hat{\sigma}_{2}^{+}+\hat{\sigma}_{2}^{-}).
\end{aligned}
\end{equation}
We cancel the frequency of the target qubit by setting $\omega_{\text{ac}} = \omega_{q_{2}} + J_{2}^{2}/(\omega_{q_{2}} - \omega_m)$. This gives Eq.~(\ref{eq:H_SW_opp}).

\end{widetext}

\bibliography{refs.bib}

\end{document}